\documentclass[preprint,aps]{revtex4}

\usepackage{graphicx}
\usepackage{dcolumn}
\usepackage{bm}
\usepackage{subfigure}

\begin{document}

\title{Numerical investigations on the finite time singularity in two-dimensional Boussinesq equations}

\author{Z. Yin}
\email{zhaohua.yin@imech.ac.cn}

\affiliation{National Micrography Laboratory, Institute of Mechanics, Chinese Academy of Sciences, Beijing 100080,
P.R.China}

\altaffiliation[Part of the work was carried out in ]{Institute of Computational Mathematics, Academy of Mathematics and
System Sciences, Chinese Academy of Sciences, Beijing 100080, P.R. China.}

\author{Tao Tang}
 \email{ttang@math.hkbu.edu.hk}
 \affiliation{Department of Mathematics, Hong Kong Baptist University, Kowloon Tong, Hong Kong, P.R. China}

\date{\today}

\begin{abstract}
To investigate the finite time singularity in three-dimensional (3D) Euler flows, the simplified model of 3D
axisymmetric incompressible fluids (i.e., two-dimensional Boussinesq approximation equations) is studied numerically.
The system describes a cap-like hot zone of fluid rising from the bottom, while the edges of the cap lag behind,
forming eye-like vortices. The hot liquid is driven by the buoyancy and meanwhile attracted by the vortices, which
leads to the singularity-forming mechanism in our simulation. In the previous 2D Boussinesq simulations, the
symmetricial initial data is used, see, e.g., \cite{e94}. However, it is observed that the adoption of symmetry leads
to coordinate singularity. Moreover, as demonstrated in this work that the locations of peak values for the vorticity
and the temperature gradient becomes far apart as $t$ approaches the predicted blow-up time. This suggests that the
symmetry assumption may be unreasonable for searching solution blow-ups. One of the main contributions of this work is
to propose an appropriate asymmetric initial condition, which avoids coordinate singularity and also makes the blow-up
to occur much earlier than that given by the previously simulations. The shorter simulation time suppresses the
development of the round-off error. On the numerical side, the pseudo-spectral method with filtering technique is
adopted. The resolutions adopted in this study vary from $1024^2$, $2048^2$, $4096^2$ to $6144^2$. With our proposed
asymmetric initial condition, it is shown that the $4096^2$ and $6144^2$ runs yield convergent results when $t$ is
fairly close to the predicted blow-up time. Moreover, as expected the locations of peak values for the vorticity and
the temperature gradient are very close to each other as $t$ approaches the predicted blow-up time.
\end{abstract}


\maketitle

\section{Introduction}

The formation of finite time singularities in the three-dimensional (3D) incompressible Euler equations is a
controversial subject in fluid mechanics. There are
two main reasons for studying the singularity development: firstly,
the verification of the finite time singularity
may aid to the understanding of the onset of turbulence in
slightly viscous flows, and secondly
if such singularities do exist then they may provide a
means by which energy cascades to and concentrates on small scales.

The equations under consideration are the following:
\begin{eqnarray}
&& {\bf u}_t + {\bf u} \cdot \nabla {\bf u} + \nabla p = 0, \label{3D-Euler-1}
\\
&& \nabla \cdot {\bf u} = 0, \label{3D-Euler-2}
\\
&& {\bf u}({\bf x}, 0) = {\bf u}_0 , \label{3D-Euler-3}
\end{eqnarray}
where, $\bf u$ is the velocity, $p$ is the pressure,
${\bf u}_0$ is the given initial velocity which is smooth in the
sense that the data reside in some Sobolov space.
However, whether the solution $\bf u$  can remain smooth for all time
has yet to be proved.

Since the question of global existence of the 3D
incompressible Euler equations has been
inaccessible analytically, the question has been investigated
numerically in the past two decades.
The most commonly used quantity in determining global existence of  Eqs.
(\ref{3D-Euler-1})-(\ref{3D-Euler-3}) is the vorticity ${\rm { \mbox{ \boldmath{$\omega$}}}} = \nabla \times {\bf u}$,
which has been first developed by Beale, Kato \& Majda ~\cite{bea84} (also see ~\cite{maj02}), and later refined by two
other groups~\cite{pon85,koz00}. It is shown in \cite{bea84}
that $\bf u$ will blow up at a finite time $T_c$ if and only if
\begin{eqnarray}
\int_0^t |{\rm { \mbox{ \boldmath{$\omega$}}}}|_{L^\infty} ds \to
\infty, \; as \;\;\;  t \nearrow T_c \;.
\label{3D-Euler-4}
\end{eqnarray}
This result is especially useful when the question of global existence is numerically investigated, because if we find
the maximum norm of the vorticity behaves like $(T_c - t)^{-\alpha}$
with $\alpha > 1$, a finite time singularity has
then developed. Based on this theory, three kinds of numerical efforts
have been  made to search for the singularities in the 3D Euler flows
\cite{sul83,bra92,ker93,tan93,bor94,pel97a,gra98,pel01,ker05a,ker05b,cic05}:
\begin{enumerate}
\item The original 3D Euler simulation does not adopt symmetric assumption.
This scheme requires the largest computer resource.
\item The symmetry introduced by Taylor \& Green~\cite{tay35}
(see also ~\cite{bra83} for some detailed discussion)
uses only $1/64$ of the total computational time for the non-symmetric case.
\item An even further symmetric technique introduced by Kida~\cite{kid85}
requires only $1/192$ of the computational time for the non-symmetric
case.
\end{enumerate}
In contrast to the 3D research, 2D study is much easier to be performed analytically or numerically. Under the
axisymmetric assumption, the 3D Euler equations can be replaced by 2D Boussinesq convection equations. This assumption
is an even more aggressive symmetric assumption than the Taylor-Green or Kida flows because it turns this 3D problem
into a 2D one, which can save computer resource significantly. We will give a brief discussion about it in Section II.
Similar to the 3D theory, the blowup judgment has been developed
for the 2D Boussinesq convection
flows~\cite{e94,cha96,cha97}. It has been realized that if
the maximum absolute values of the vorticity and temperature
gradient behave like $(T_c - t)^{-\alpha}$
and $(T_c - t)^{-\beta}$ with $\alpha > 1$ and $\beta > 2$, a finite time
singularity will be developed. This simplified 2D model, although
lacking of the vortex reconnection as in the 3D
simulations~\cite{ash87,ker89,mof00},
can also reveal certain possibility of
singularity formation ~\cite{e94,gra91,pum90,pum92a,pum92b,gra95,car97,cen01}.

Among the numerical efforts in addressing the singularity issue,
some highest possible resolutions have been attempted in the last two decades.
There are mainly two kinds of numerical schemes involved: pseudo-spectral
methods and finite differences.
For the pseudo-spectral efforts, Kerr used the resolution up
to $1024 \times 256 \times 128$ without any symmetric
assumption~\cite{ker93}, and in a recent work of Hou and Li ~\cite{hou06}
an extremely fine grid of resolution $1536 \times 1024 \times 3072$ is used.
The Taylor-Green or Kida simulations have reached the effective
resolution of $1024^3$~\cite{bor94} and
$2048^3$~\cite{cic05}.  Moreover, the most intensive axisymmetric
simulation~\cite{e94} uses a resolution of $1500^2$.

To the best of out knowledge, the most intensive 3D finite difference
calculation was performed by Grauer \emph{et
al.}~\cite{gra98}. They did not make any symmetric
assumption, but used the adaptive mesh refinement (AMR)
method to enhance the effective resolution to $2048^3$. For
the 2D Boussinesq simulation, uniform grids of $512^2$ and $1792 \times 1280$
were used in ~\cite{e94} and ~\cite{gra95}, respectively.
Pumir \& Siggia ~\cite{pum92a} employed an adaptive $256^2$ grid to
achieve a resolution of $10^7$ in both dimensions.
A less aggressive adaptive grid, maybe more accurate due to
less frequent re-meshing,
uses a $512^2$ deformed grid to reach a $4600^2$ effective
resolution~\cite{cen01}.

Besides the numerical efforts mentioned above, the 3D
Taylor series analysis, although limited due to the lack
of proper parallelized softwares handling high
precision calculations, is also used to analyze the 3D Euler
singularity problem~\cite{mei82,pel97b,pel03,gul05}. Moreover,
some analytical studies have been carried
out~\cite{moo79,bha92,bha95,ng96,gre97,gre00,rub01,ohk00,gib01,gib03},
but the original 3D Euler
equations have been modified which makes the use of relevant
mathematical tools possible,
Likewise, some 2D modified Boussinesq equations were also employed
in simulations and analysis \cite{san89,con94,maj96,mcw98}.
Although these models do not have much connection
with the original 3D Euler equations, they may provide some hints
to the understanding of the singularity issue. One of the latest results
was given by Frisch \emph{et al.} ~\cite{fri03,mat05} who conducted the
2D calculations in the so-called parareal domain by taking
the advantages of both spectral and adaptive methods.

One of the main purposes of this work is to investigate
how to set up an effective initial condition which can be used
in the simulations of the Boussinesq equations.
The reason that we emphasize the importance of proper initial conditions
is that a poorly chosen initial condition may not
lead to blowup or may lead to blowup at a numerically unacceptable large time.

Section II contains a brief discussion on our numerical method.
We adopted the so-called phase-shifted technique to do
the de-aliasing~\cite{ors71,can87}.
It is clear that to a reliable and efficient
numerical is very important in simulating possible singular behaviors.
Section III gives a quite complete discussion of three initial conditions,
including subtle but significant differences from earlier work.
In section V, we use the parallel strategy combined with the
traditional parallel FFT and task distribution schemes~\cite{yin04,yin05}
to solve the 2D Boussinesq equations
with resolutions up to $6144^2$. The corresponding effective
resolution in the 3D Euler formulation is $\pi \times
6144^3 \simeq 9000^3$, which is much finer than any previous
efforts. Singularity development will be demonstrated by
considering several physical quantities including the
peak vorticity and temperature gradient.

\section{The numerical scheme}

\begin{figure}
\begin{minipage}[c]{.6 \linewidth}
\scalebox{1}[1]{\includegraphics[width=\linewidth]{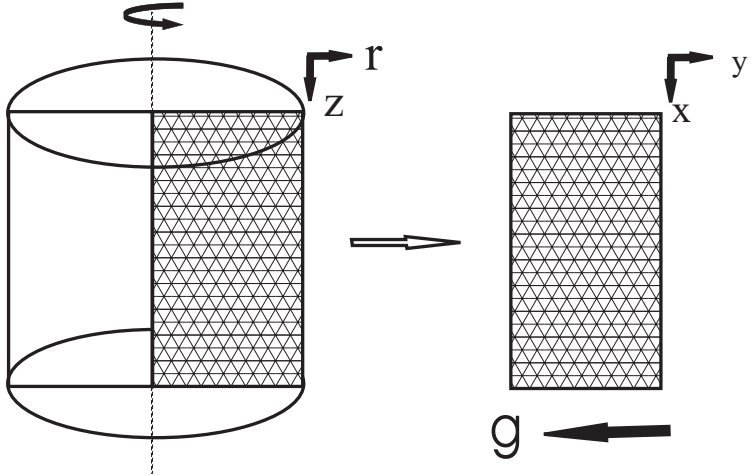}}
\end{minipage}
\caption{\label{fig:2_3D} The analogy between the $(r, z)$ plane in 3D axisymmetric flows with swirl and the $(x, y)$
plane of 2D Boussinesq convection flows. We keep $\textbf{u} \equiv 0$ near $y = 0$ and $y = 2\pi$ throughout our 2D
Boussinesq doubly-periodic calculation to avoid the coordinate singularity in the corresponding 3D axisymmetric flows
(see Fig. \ref{fig:3_initial}(a)). The solid black arrow at the bottom indicates the direction of the gravity, opposite
to the direction of buoyancy.}
\end{figure}

The connection between the 3D axisymmetric Euler flow with swirl and the
2D Boussinesq convection has been established in
~\cite{maj02}. For completeness, we will review the connection briefly.

The basic vorticity equations for the
axisymmetric swirling flow in the $(r,\theta,z)$ cylindrical coordinates are
of the form
\begin{eqnarray}
&& \frac{\tilde{D}}{Dt}(r v^{\theta})=0, \label{cylindrical-1}
\\
&& \frac{\tilde{D}}{Dt}({\frac{\omega^{\theta}}{r}})=- \frac{1}{r^4} [{(r v^{\theta})^2}]_z ,  \label{cylindrical-2}
\end{eqnarray}
where
\begin{eqnarray}
&& \frac{\tilde{D}}{Dt}=\frac{\partial}{\partial t} + v^r \frac{\partial}{\partial r} + v^z \frac{\partial}{\partial z},
\label{cylindrical-3}
\\
&& \omega^{\theta} = \frac{\partial v^r}{\partial z} - \frac{\partial v^z}{\partial r}.  \label{cylindrical-4}
\end{eqnarray}
It should be realized that the $(z,r)$ plane is essentially two
dimensional flows with $(x,y)$ coordinate system (Fig.
\ref{fig:2_3D}). By introducing a streamfunction $\psi$, the
2D inviscid Boussinesq equations have the following form:
\begin{eqnarray}
&& \theta_t + {\rm {\bf u}} \cdot \nabla \theta = 0, \label{boussinesq-1}
\\
&& \omega _t + {\rm {\bf u}} \cdot \nabla \omega = - \theta_x, \label{boussinesq-2}
\\
&& \Delta \psi = - \omega, \label{boussinesq-3}
\end{eqnarray}
where the gravitational constant is normalized to $\textbf{g}=(0,-1)$,
$\theta$ the temperature, ${\rm {\bf u}} =
(\mbox{u,v})$ the velocity, ${\rm { \mbox{ \boldmath{$\omega$}}}}
= (0,0,\omega ) = \nabla \times {\rm {\bf u}}$
vorticity, and $\psi$ the stream function.
It should be pointed out in the remaining of the work
 $\theta$ is different with
that indicated in (\ref{cylindrical-1})-(\ref{cylindrical-2});
it denotes the temperature.
Comparing with the pure 2D Euler equations with the streamfunction-vorticity
$\omega$-$\psi$ formulation,
there is an extra temperature equation, i.e. Eq. (\ref{boussinesq-1}),
and an extra term $\theta_x$ associated with the
Buoyancy in Eq. (\ref{boussinesq-2}). Actually, if
$\omega^{\theta}$ in Eqs. (\ref{cylindrical-1}-\ref{cylindrical-4})
is replaced by $\omega$, and $(r v^{\theta})^2$ by $\rho$,
a link can be established if we evaluate all external variable
coefficients in Eq. (\ref{cylindrical-1}-\ref{cylindrical-4})
at $r = 1$. It should be
noticed that this link leads to some coordinate singularities under
the cylindrical coordinates. Consequently, it is
essential to keep the corresponding 2D Boussinesq solutions
away from the horizontal boundaries (here, in this paper, $y
= 0$ and $y = 2 \pi$; see Figs. \ref{fig:2_3D} and Fig. \ref{fig:3_initial}(a)).

The method used in our numerical simulation is pseudo-spectral approximations
with some proper de-aliasing technique. We also
adopt the filtering technique ~\cite{van91} (a careful discussion on filter
in turbulence simulations can be found in \cite{shu05})
to modify the Fourier coefficients such that the stability of the
numerical scheme is enhanced. The machine
accuracy of our computer with double precision is $\epsilon = 10^{-16} \approx e^{-37}$, and the modifying factor in the
filter is $\varphi(k) = e^{-37(2k/N)^{16}}$ for $k < N/2$, where
$N$ is the Fourier modes in each direction.

We now briefly discuss the effects of different
numerical schemes for the current research. As to be shown later,
the main concern is
whether or not a $\delta$ or $\delta$-like function can be well resolved
by using a discrete scheme. It is well
known that the spectral coefficients of the $\delta$ function is
constant for different modes:
\begin{equation}
\delta(x,y) \simeq C \sum^{\frac{N}{2}}_{n = -\frac{N}{2}} \; \sum^{\frac{N}{2}}_{m = -\frac{N}{2}} e^{-i(nx + my)},
\label{delta-1}
\end{equation}
where $C$ is a constant and will be set to $1$ in the following. If
the resolution $N^2 \rightarrow \infty$, we will get
the exact Fourier representation. Of course, it is impossible to do this by
any computer, and what we can do is to use some finest possible
grid (the largest $N^2$
appearing in the literature is limited to $8192^2$ so far~\cite{dmi05}). The
de-aliasing technique is to remove the aliasing error by setting
some of the Fourier coefficients to zero. The filter we
adopted is to make some of the Fourier coefficients smaller.
So actually, the computer-represented value of the
$\delta$ function is determined by the resolution $N^2$, the filter and
the de-aliasing.

\begin{table}
\caption{\label{table1}Peak values when a $\delta(x,y)$ function is represented by different schemes with the same
resolution $N^2$.} \centering
\begin{tabular}
{|p{70pt}|p{105pt}|p{105pt}|p{105pt}|} \hline \hline
&peak value before filtering & percentage retained by the filter & peak value after filtering \\
\hline phase-shift & $\;2\pi N^2 /9$ & $\;81.3 \%$ & $\;0.568 N^2$ \\
\hline No de-aliasing & $\;N^2$& $\;59.6\% $ & $\;0.596 N^2$ \\
\hline \hline
\end{tabular}
\end{table}

The de-aliasing scheme to be used is the so-called phase-shift
scheme~\cite{can87}, which retains about $7/9$ of the
total modes. The filter narrows the gap of the peak values of
de-aliasing pseudo-spectral schemes and pure spectral
schemes. In Table I, the filtering $\delta$ function value of phase-shift scheme is only $2 \%$ lower than the no
de-aliasing scheme. In our calculation, we follow the tradition of our pseudo-spectral code (W. H. Matthaeus, private
communication) to use the circular truncation in our running~\cite{pat71}.

\section{Initial conditions}

In this paper, three different initial conditions will be
considered, which will be named RUN A, RUN B and
RUN C, respectively.
RUN A adopted the same initial condition as used
in Ref ~\cite{e94}:
\begin{eqnarray}
&& \omega (x,y,0) = 0, \label{ec94} \\
\label{eq33}
&& \theta(x,y,0) = 50\theta _1 (x,y)\theta _2 (x,y)
\left[ {1 - \theta _1 (x,y)} \right], \label{eq33}
\end{eqnarray}
where if $S(x, y):=\pi^2 - y^2 - (x - \pi )^2$ is positive, $\theta_1 = \exp {\left( 1 - \pi^2/S(x,y)\right)}$, and
zero otherwise; if $s(y):= \left| {y - 2\pi } \right| /1.95\pi$ is less than 1, $\theta_2 = \exp \left( 1- (1- s(y)^2)
^{-1} \right)$, and zero otherwise. By choosing the initial conditions (\ref{ec94})-(\ref{eq33}), we can test our
discretization schemes by comparing our numerical results with those given in \cite{e94}.

\begin{figure*}
\begin{minipage}[c]{.29 \linewidth}
\scalebox{1}[1.]{\includegraphics[width=\linewidth]{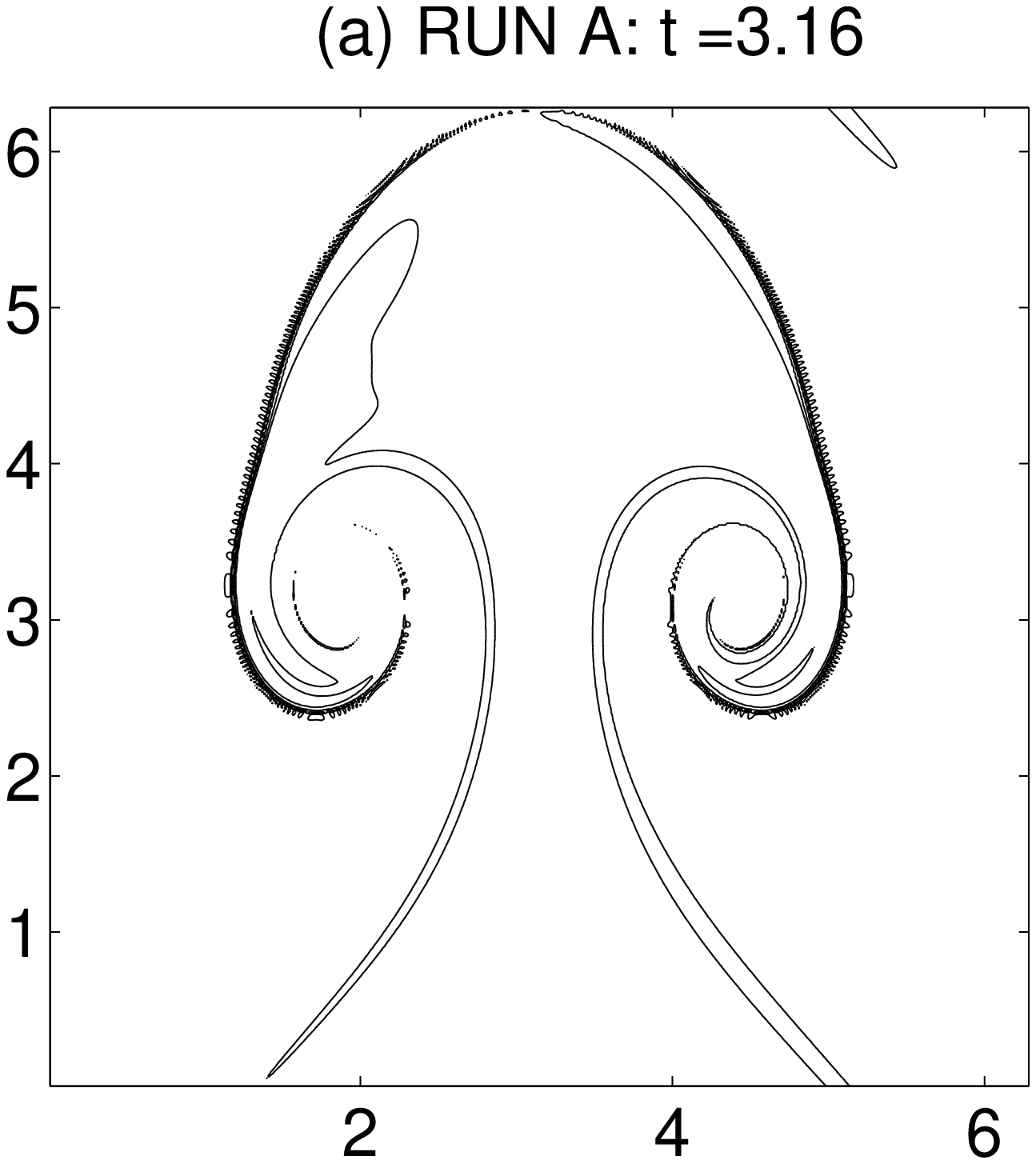}}
\end{minipage}
\begin{minipage}[c]{.29 \linewidth}
\scalebox{1}[1.]{\includegraphics[width=\linewidth]{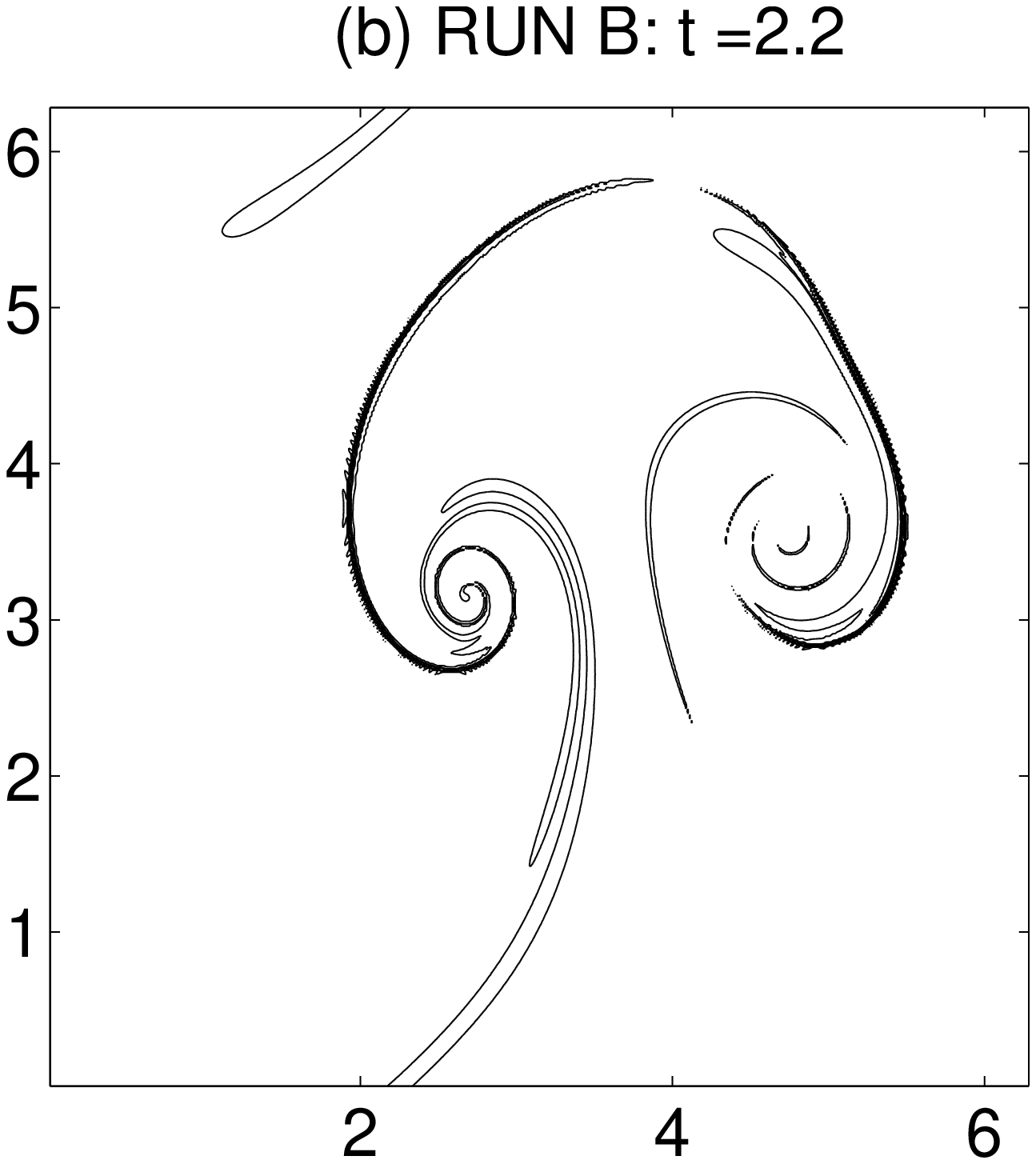}}
\end{minipage}
\begin{minipage}[c]{.29 \linewidth}
\scalebox{1}[1.]{\includegraphics[width=\linewidth]{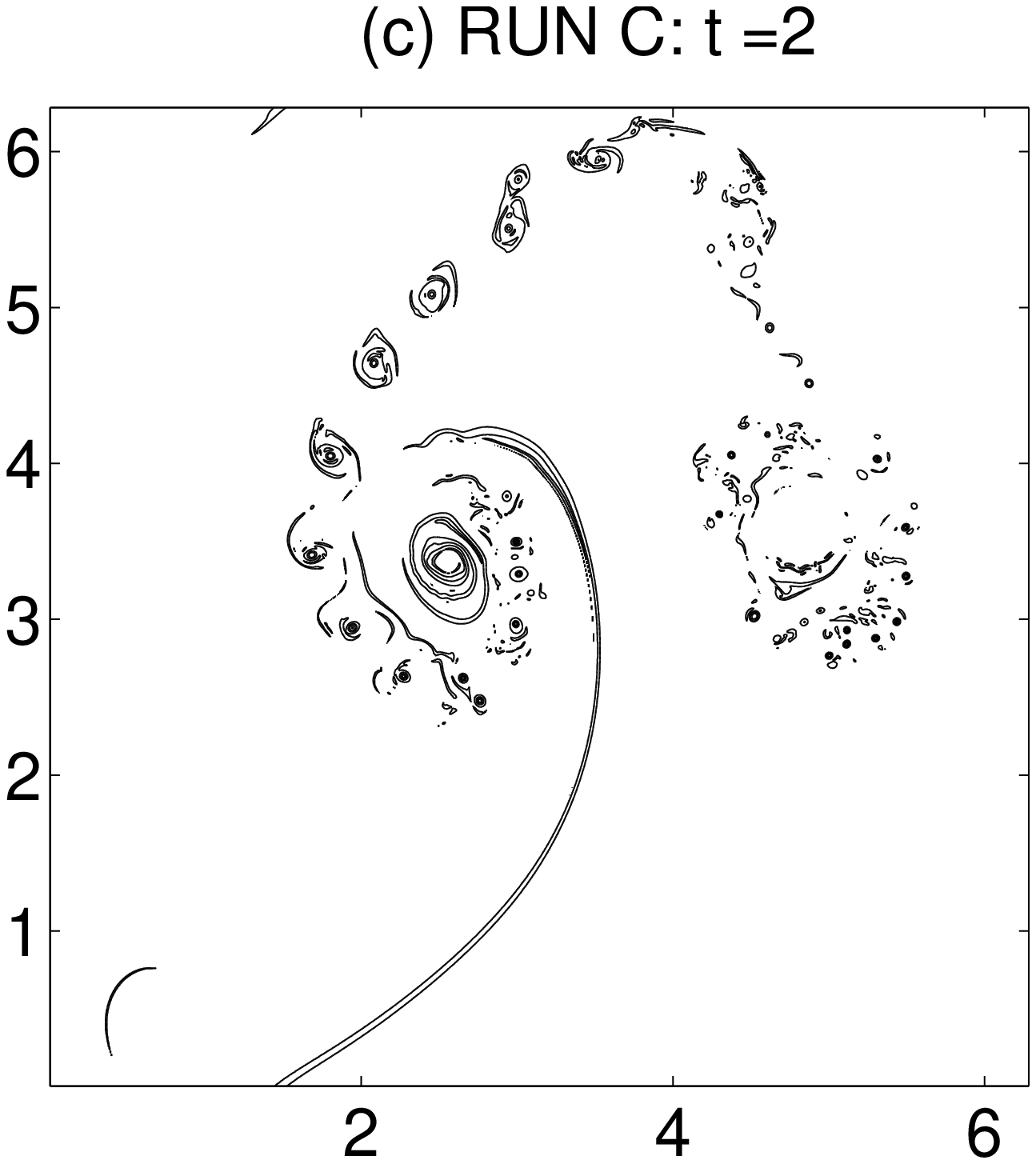}}
\end{minipage}
\caption{\label{fig:3_cross} The vorticity contours of three runs when non-velocity crosses the horizontal boundary. a)
and b) are before the blow-up time for RUN A and B, and c) is after the blow-up time of RUN C.}
\end{figure*}

\begin{figure}
\begin{minipage}[c]{.45 \linewidth}
\scalebox{1}[1]{\includegraphics[width=\linewidth]{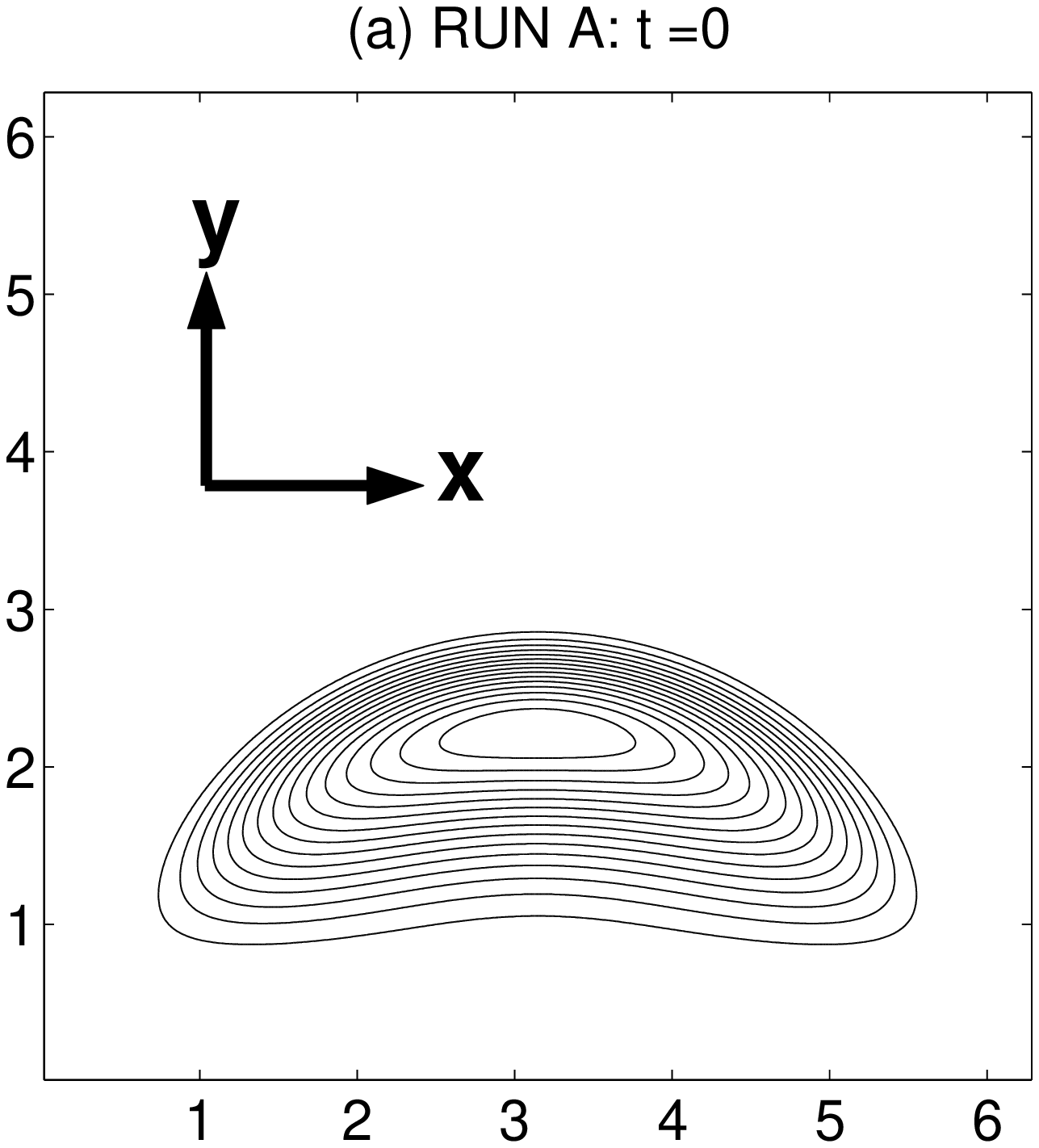}}
\end{minipage}
\begin{minipage}[c]{.45 \linewidth}
\scalebox{1}[1]{\includegraphics[width=\linewidth]{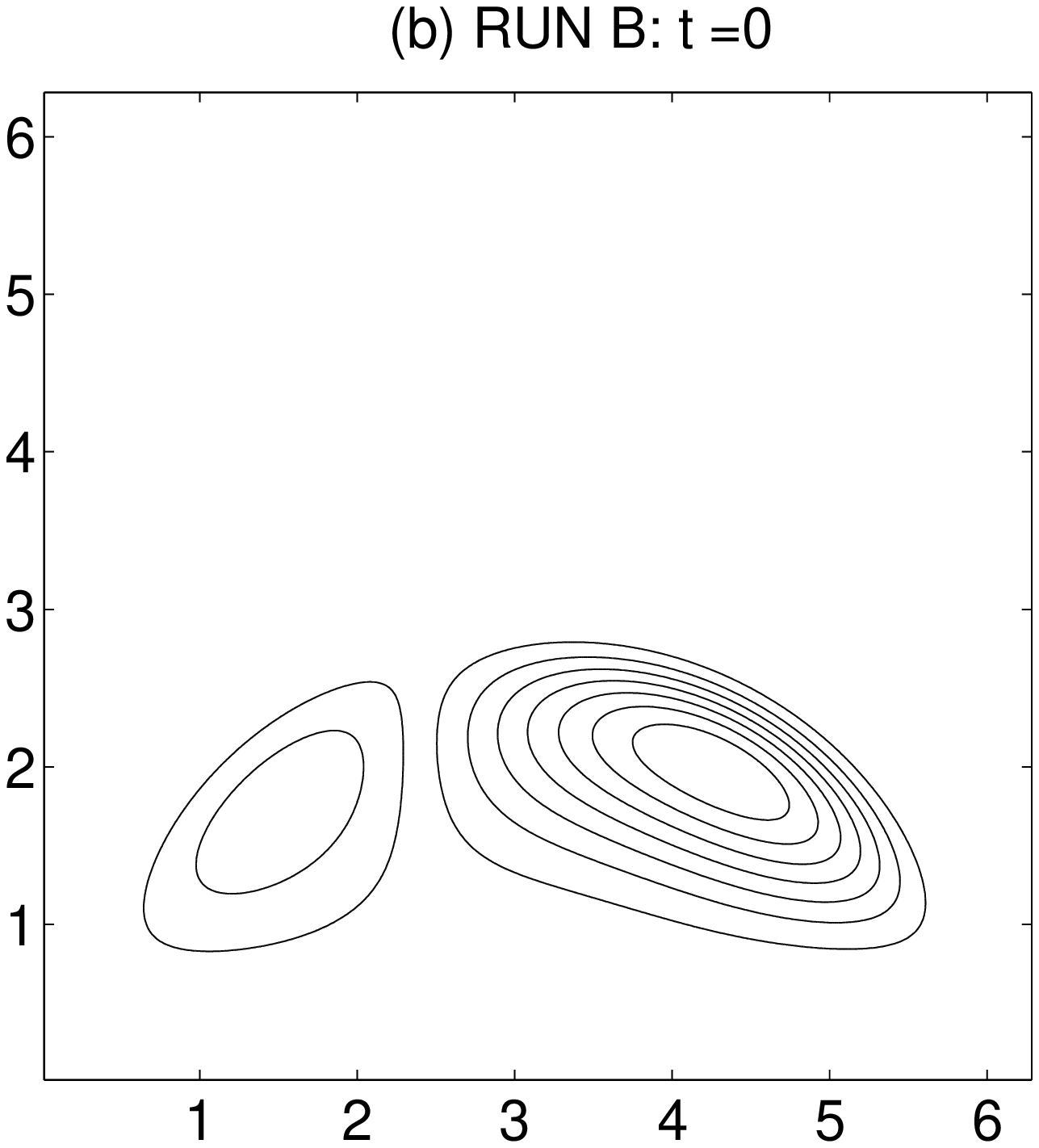}}
\end{minipage}
\begin{minipage}[c]{.45 \linewidth}
\scalebox{1}[1]{\includegraphics[width=\linewidth]{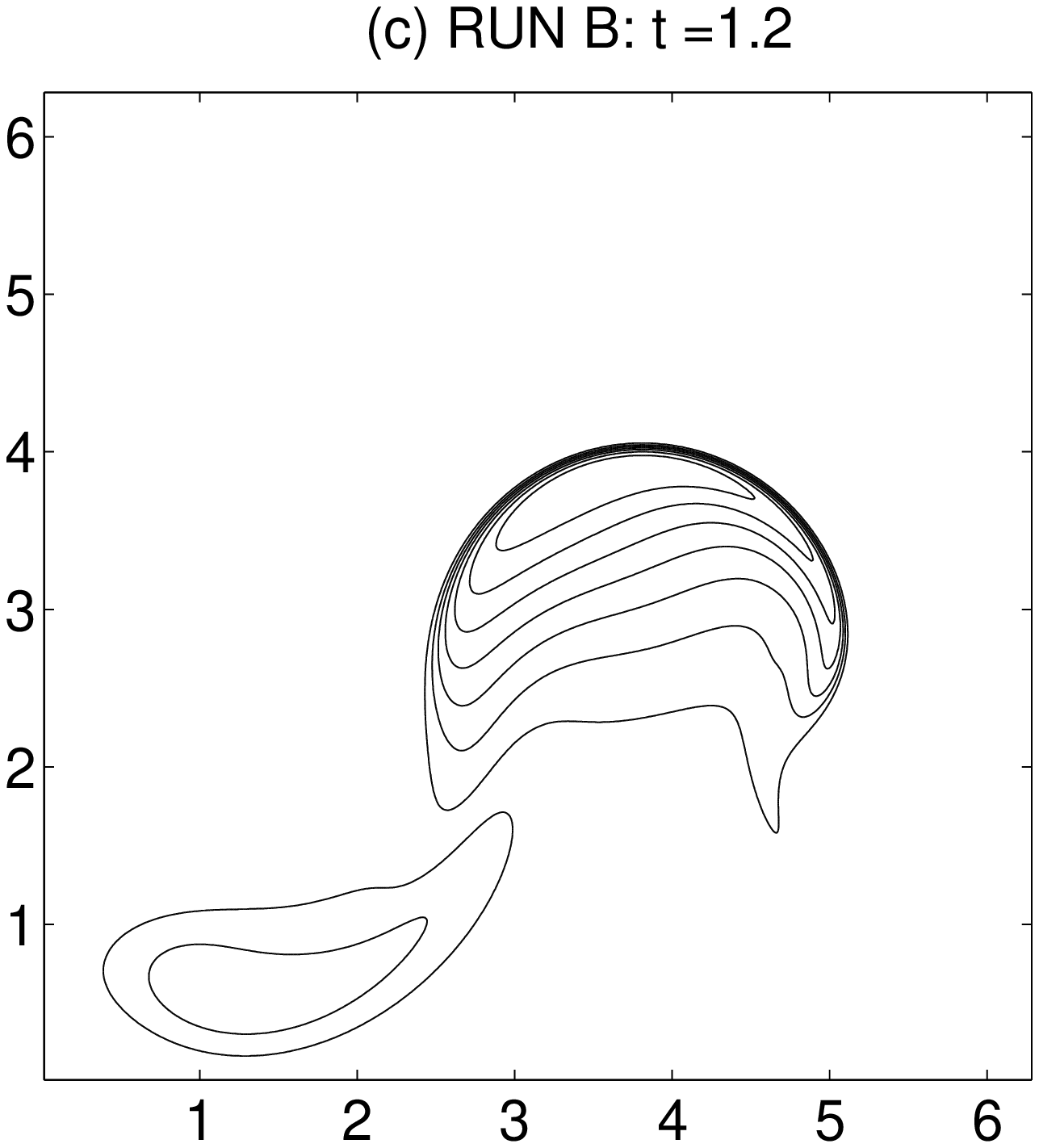}}
\end{minipage}
\begin{minipage}[c]{.45 \linewidth}
\scalebox{1}[1]{\includegraphics[width=\linewidth]{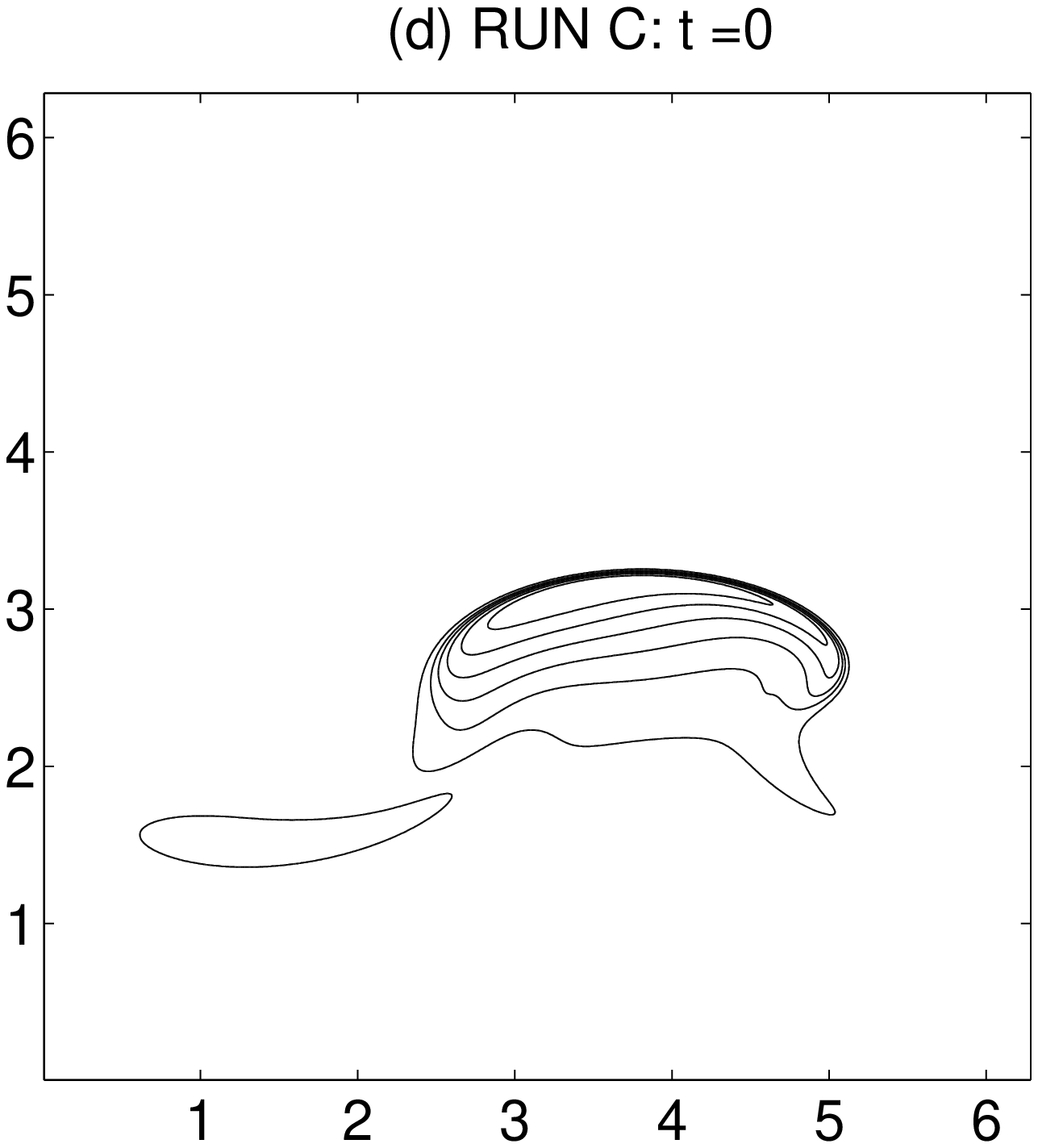}}
\end{minipage}
\caption{\label{fig:3_initial}a), b) and d) are contour plots of temperature for three different initial conditions. d)
is obtained after the intermediate results of RUN B (c) are compressed by the factor of 2.}
\end{figure}

This governing equations (\ref{boussinesq-1})-(\ref{boussinesq-3})
with the above initial condition have been studied intensively
by E \& Shu ~\cite{e94} with spectral methods (resolutions:
$1500^2$) and ENO finite difference methods
(resolutions: $512^2$). They have predicted that the side part of the
rising bubble is much more dangerous than the front of the bubble (Fig. \ref{fig:3_cross}(a)). However, this initial
condition has two shortcomings:
\begin{itemize}
\item It introduces another symmetry assumption
besides the axisymmetric assumption described in the previous section: the
symmetry respect to $x = \pi$ in the $(x, y)$ plane. This will cause a further deviation from the original 3D non-symmetric Euler flows.
\item It is demonstrated that the blow-up time of this flow is after $t =3.16$,
at which both $y = 0$ and $y = 2 \pi$ are crossed by non-zero
velocity, see Fig. \ref{fig:3_cross}(a). This is due to the assumption
we adopted to simplify the 3D axisymmetric flow to the 2D
Boussinesq flow. In other words, what we are studying here can not be
closely related to the 3D Euler flow anymore.
\end{itemize}

To fix the first difficulty above, we change the initial temperature
field in (\ref{eq33}) to
\begin{equation}
\label{ec94as} \theta (x,y,0) = \frac{50}{\pi}(4x-3\pi)
\theta _1 (x,y)\theta _2 (x,y)\left[ {1 - \theta _1 (x,y)}
\right].
\end{equation}
The factor $(4x-3\pi)/\pi$ introduced here is to break the symmetry
assumption with respect to $x = \pi$ in the $(x, y)$ plane.
However, this simulation
(denoted as RUN B) will also cross both $y = 0$ and $y = 2 \pi$
(see Fig. \ref{fig:3_cross}(b)) before $T_c$. Therefore, a better
initial condition (RUN C) is to
compress the intermediate results at $t=1.2$ obtained in RUN B. More
precisely, we let
\begin{equation}
\omega(x,y,0) = \omega'(x,2y-0.4\pi,1.2), \quad
\theta(x,y,0) = \theta'(x,2y-0.4\pi,1.2),
\end{equation}
for $(x,y)\in [0, 2\pi] \times [0, \pi]$
(where $\theta'$ and $\omega'$ are obtained by solving Eqs.
(\ref{boussinesq-1})-(\ref{boussinesq-3}),
(\ref{ec94}) and (\ref{ec94as}) with a $2048^2$ grid), and zero otherwise.
The initial condition associated with RUN C is demonstrated
in Figs. \ref{fig:3_initial}(b, c, d).

There are three main advantages for using the above initial conditions
with RUN C:
\begin{itemize}
\item There will have no symmetric and boundary crossing problems
as observed in RUN A. The flow pattern does not cross the
horizontal boundary until $t = 2$ (Fig. \ref{fig:3_cross}(c))
while the predicted blow-up time is around $t = 0.91$
(see Section IV B).

\item Compared with RUN A, RUN C only needs about 1/4 simulation
time to reach the blow-up time $T_c$, and the advantages of higher
resolutions show up at early stages of the simulations.
As a result, RUN C can save quite large amount
of computational time, which is particularly important
in this kind of study.

\item Shorter simulation time also suppresses the development of
round-off error, which is non-trivial in the current
high resolution simulations (see Appendix).
\end{itemize}

\section{Numerical results}
In this section we will present a detail discussion of the numerical results of RUN A and RUN C. Some features of Run
B can be represented by RUN A and C, and will be only briefly discussed.
RUN A will serve as a validate run. Moreover, it is used
to illustrate some pitfalls we should avoid,
such as round-off error, symmetric effect, and coordinate
singularity etc.
RUN C is the simulation from which we will draw main conclusions.

\subsection{Numerical results for RUN A}

\begin{figure*}
\begin{minipage}[c]{.32 \linewidth} 
\scalebox{1}[1.3]{\includegraphics[width=\linewidth]{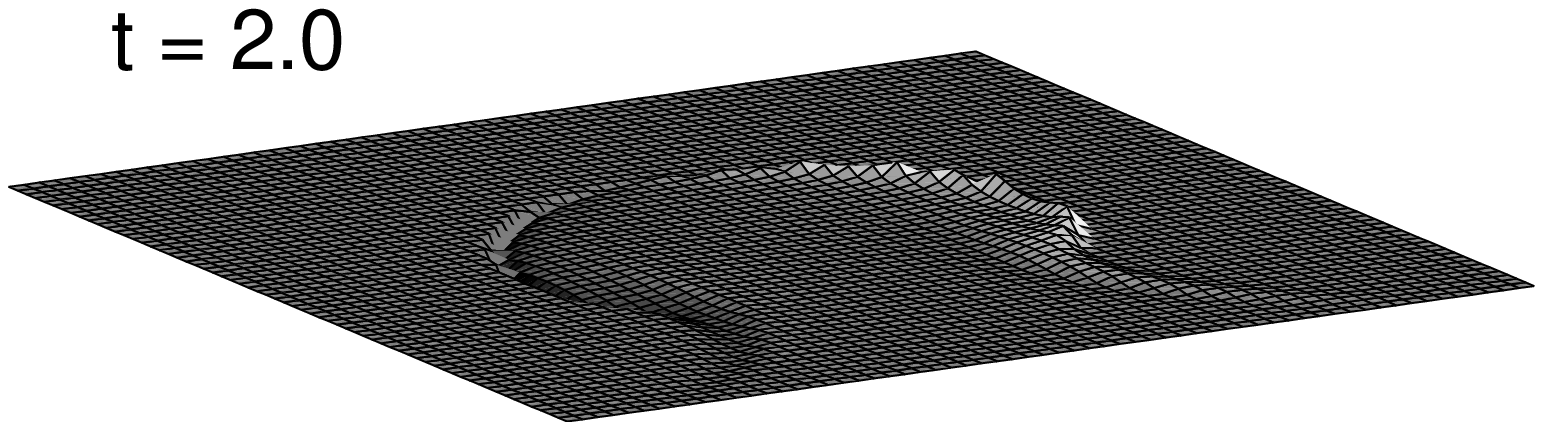}}
\end{minipage} 
\begin{minipage}[c]{.32 \linewidth}
\scalebox{1}[1.3]{\includegraphics[width=\linewidth]{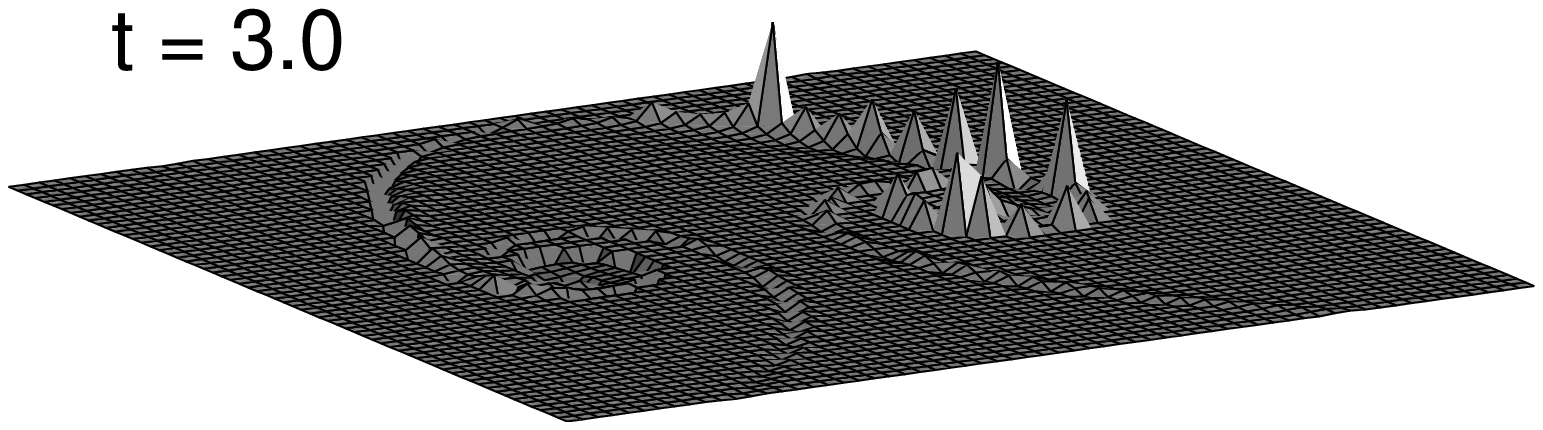}}
\end{minipage} 
\begin{minipage}[c]{.32 \linewidth} 
\scalebox{1}[1.3]{\includegraphics[width=\linewidth]{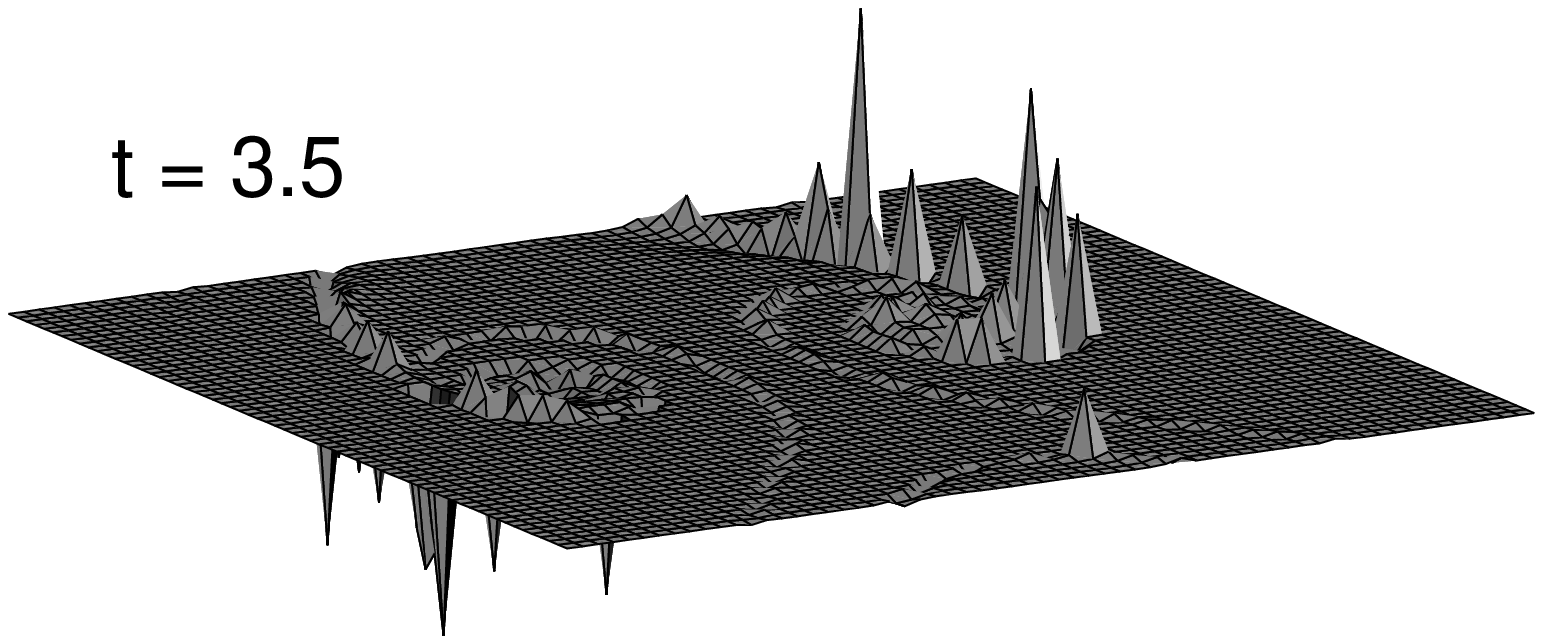}}
\end{minipage} 
\caption{\label{fig:A3dvor} The 3D perspective plots of vorticity for RUN A with a $4096^2$ grid at different times.
All 3D effect pictures in this paper (Figs. \ref{fig:A3dvor}, \ref{fig:A3dp} and \ref{fig:C3dvor}) are based on smooth
flow field.}
\end{figure*}

\begin{figure*}
\begin{minipage}[c]{.32 \linewidth}
\scalebox{1}[0.95]{\includegraphics[width=\linewidth]{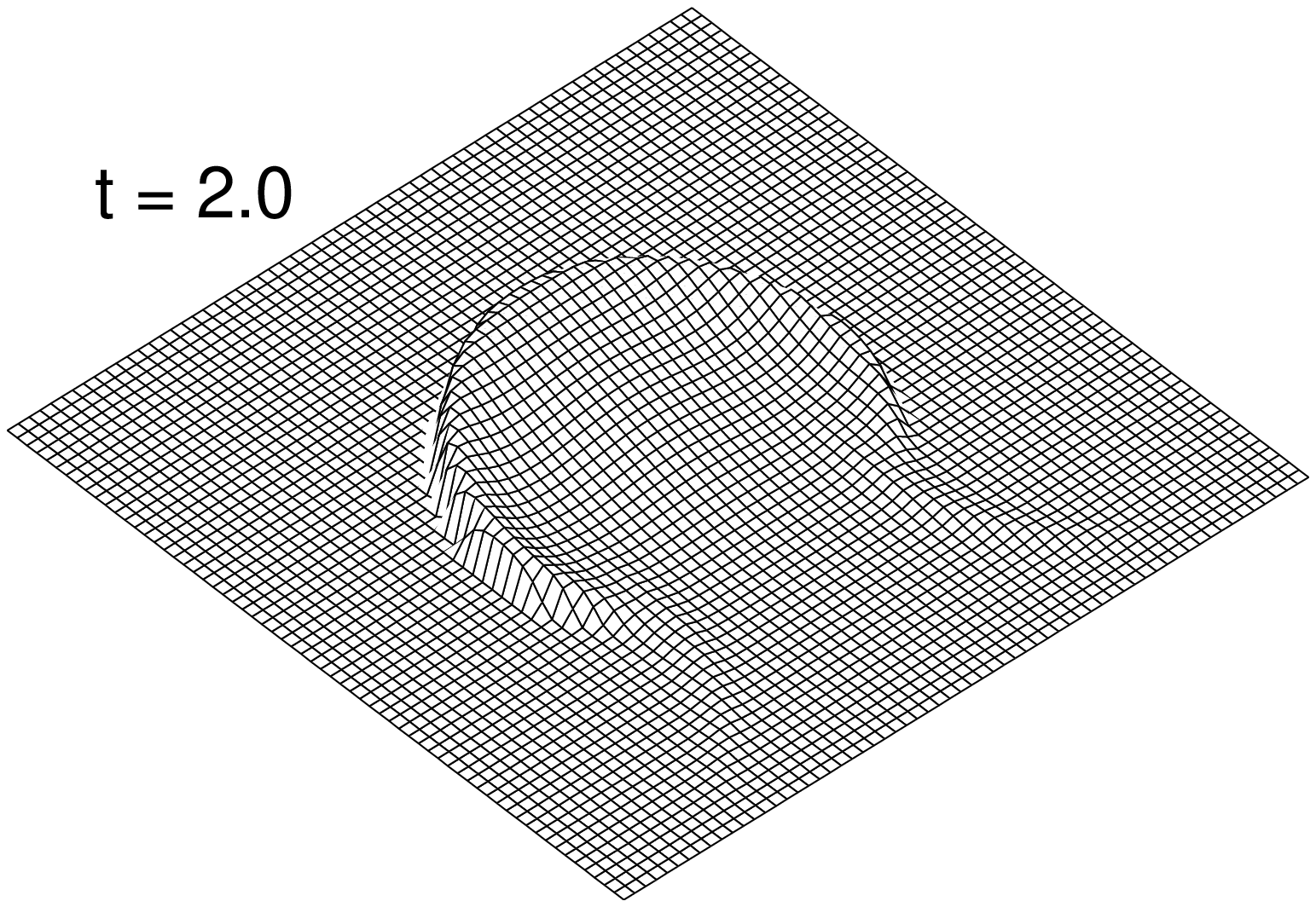}}
\end{minipage}
\begin{minipage}[c]{.32 \linewidth}
\scalebox{1}[0.95]{\includegraphics[width=\linewidth]{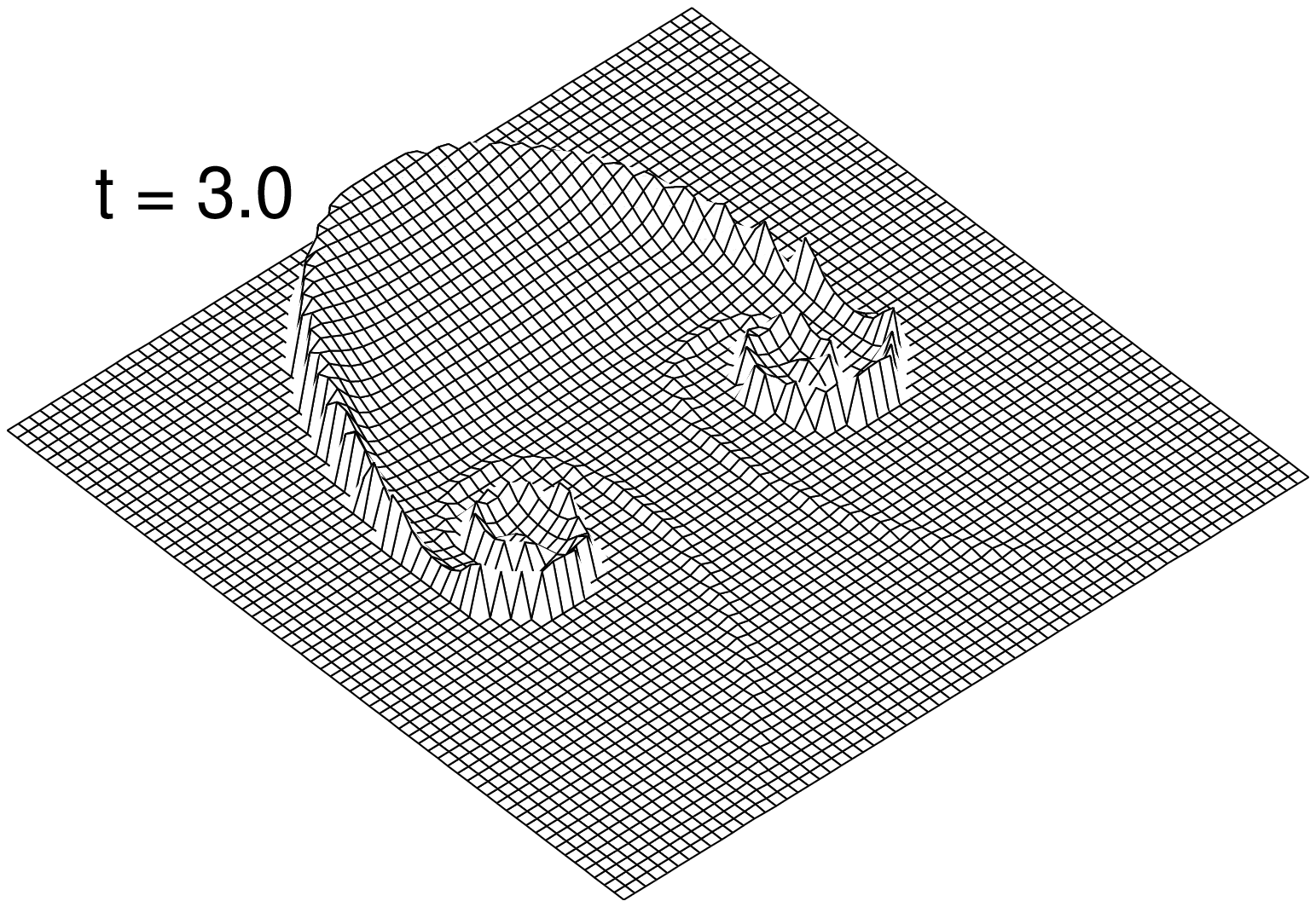}}
\end{minipage}
\begin{minipage}[c]{.32 \linewidth}
\scalebox{1}[0.95]{\includegraphics[width=\linewidth]{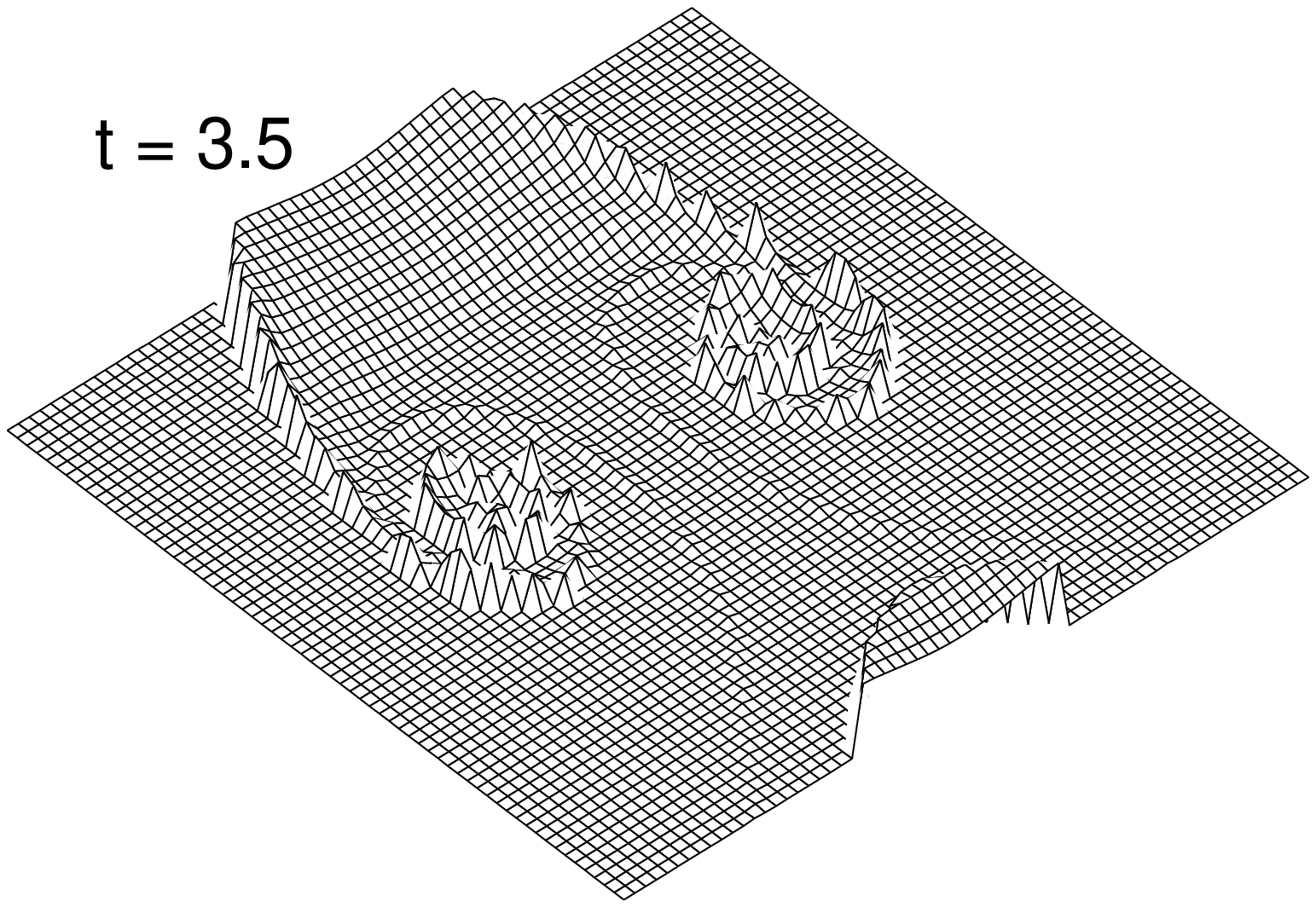}}
\end{minipage}
\caption{\label{fig:A3dp} The 3D perspective plots of temperature for RUN A
 with a $4096^2$ grid.}
\end{figure*}

\begin{figure*}
\begin{minipage}[c]{.9 \linewidth}
\scalebox{1}[1]{\includegraphics[width=\linewidth]{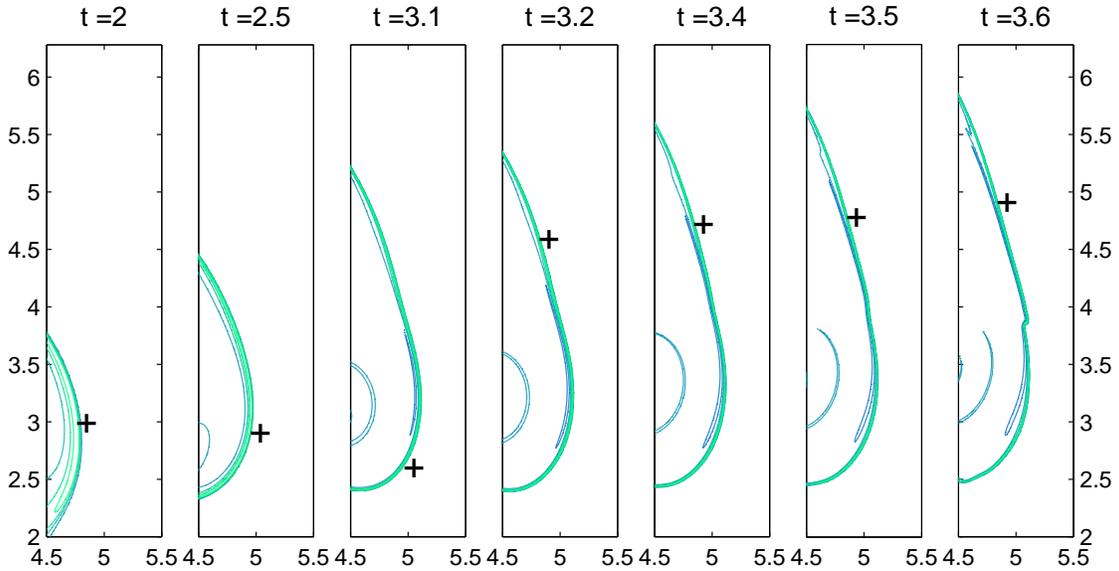}}
\end{minipage}
\caption{\label{fig:Azoomv} (Color online). Zoom-in contour plots of vorticity in RUN A at different times with the
resolution of $4096^2$. Only details near the predicted singularity location ($[4.5,5.5] \times [2, 2 \pi]$) are shown.
The plus symbol indicates the location of $|\omega|_{max}$ in the whole $[0, 2 \pi] \times [0, 2 \pi]$ domain.}
\end{figure*}

\begin{figure*}
\begin{minipage}[c]{.9\linewidth}
\scalebox{1}[1]{\includegraphics[width=\linewidth]{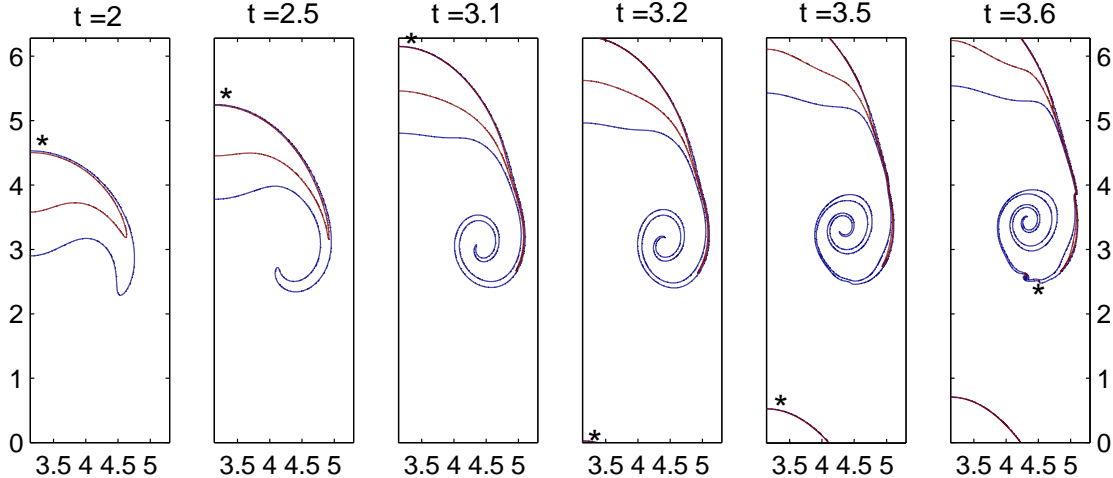}}
\end{minipage}
\caption{\label{fig:Azoomp} (Color online). Zoom-in contour plots of temperature in RUN A at different times with the
resolution of $4096^2$. Only details in $[3.25,5.3] \times [0, 2 \pi]$ are shown. The ``$*$'' symbol indicates the
location of $|\nabla \theta|_{max}$ in the whole $[0, 2 \pi] \times [0, 2 \pi]$ domain.}
\end{figure*}

\begin{figure*}
\begin{minipage}[c]{.495 \linewidth}
\scalebox{1}[1.2]{\includegraphics[width=\linewidth]{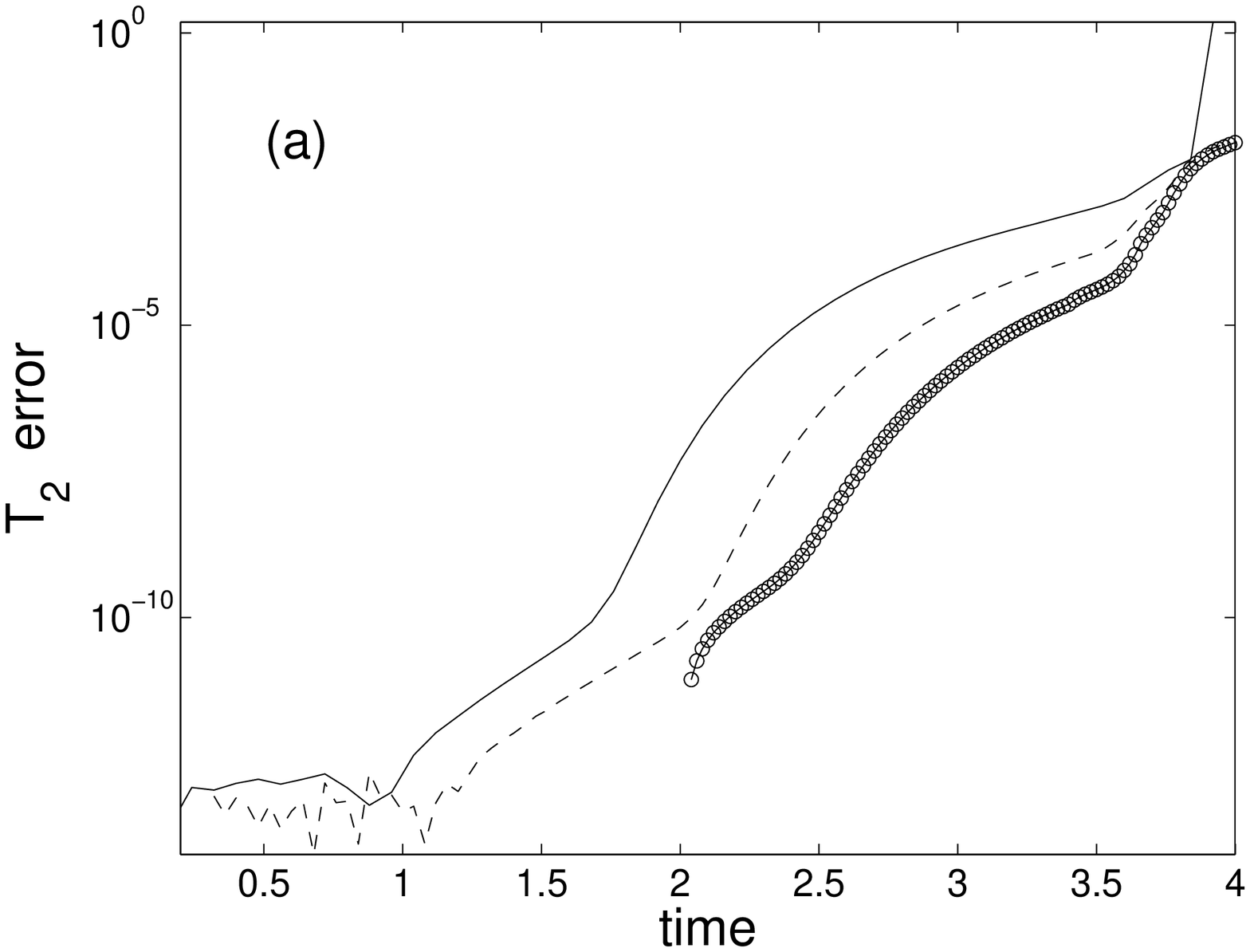}}
\end{minipage}
\begin{minipage}[c]{.495 \linewidth}
\scalebox{1}[1.2]{\includegraphics[width=\linewidth]{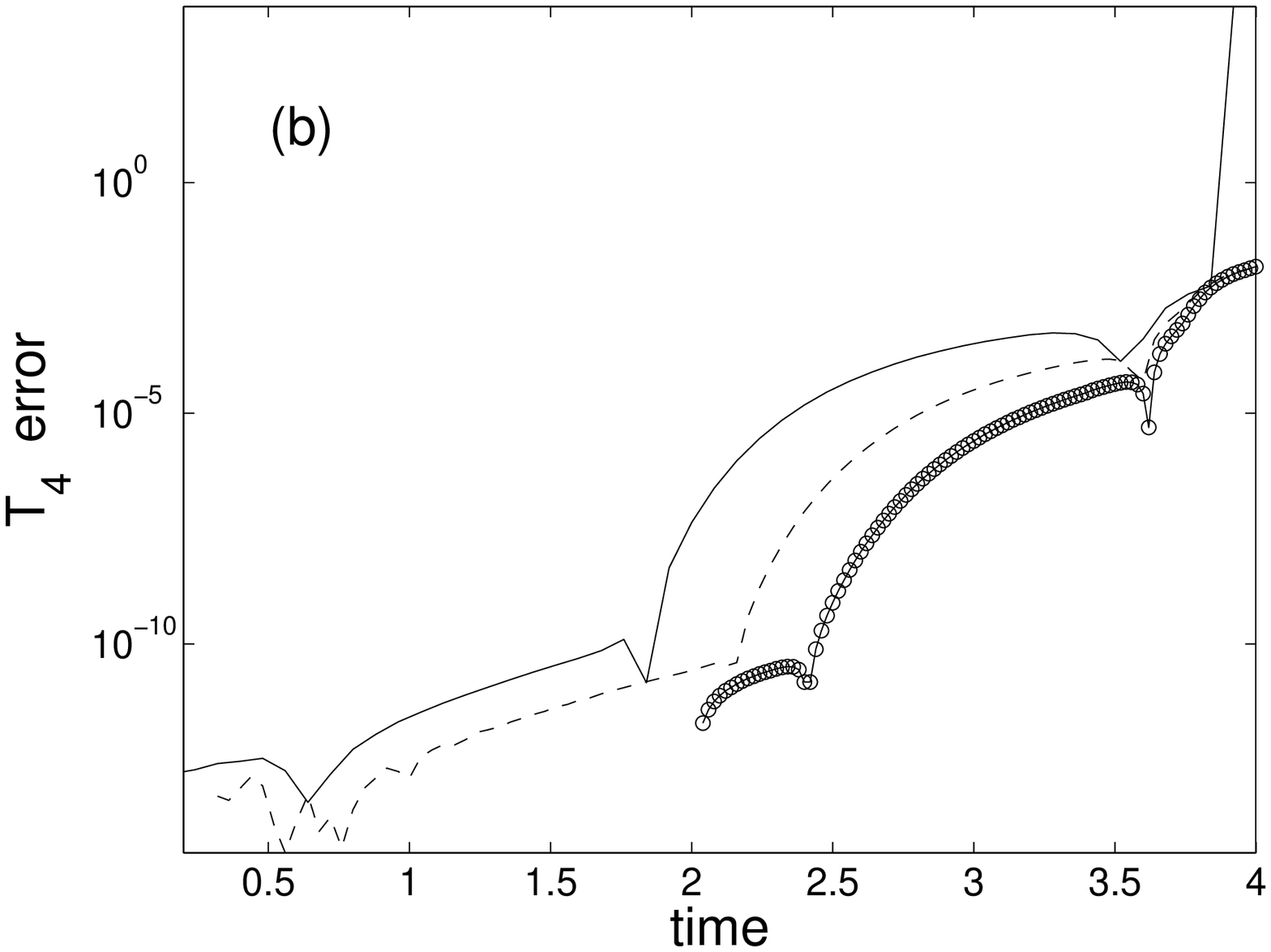}}
\end{minipage}
\begin{minipage}[c]{.495 \linewidth}
\scalebox{1}[1.2]{\includegraphics[width=\linewidth]{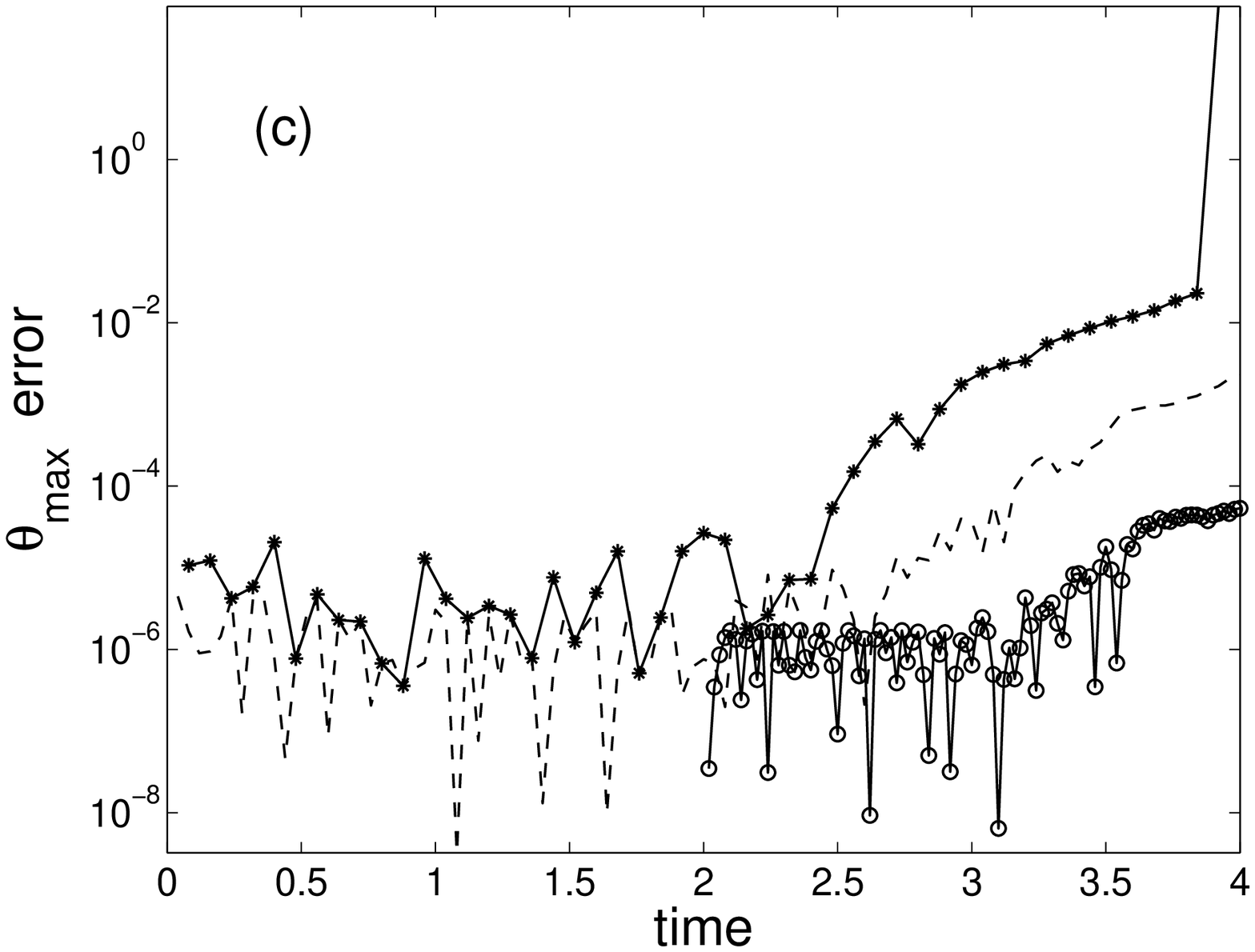}}
\end{minipage}
\begin{minipage}[c]{.495 \linewidth}
\begin{minipage}[c]{.99 \linewidth}
\caption{\label{fig:Aerror} RUN A: Time evolutions of the relative errors in $T_2$, $T_4$ and $\theta_{max}$ (solid
line: $1024^2$; dash line: $2048^2$; circle line: $4096^2$), defined as ${(T_2(0)-T_2(t))}/{T_2(0)}$,
${(T_4(0)-T_4(t))}/{T_4(0)}$ and $|\theta_{max}(0) - \theta_{max}(t)|/|\theta_{max}(0)|$. }
\end{minipage}
\end{minipage}
\end{figure*}

\begin{figure*}
\begin{minipage}[c]{.8 \linewidth}
\begin{minipage}[c]{.489 \linewidth}
\scalebox{1}[1]{\includegraphics[width=\linewidth]{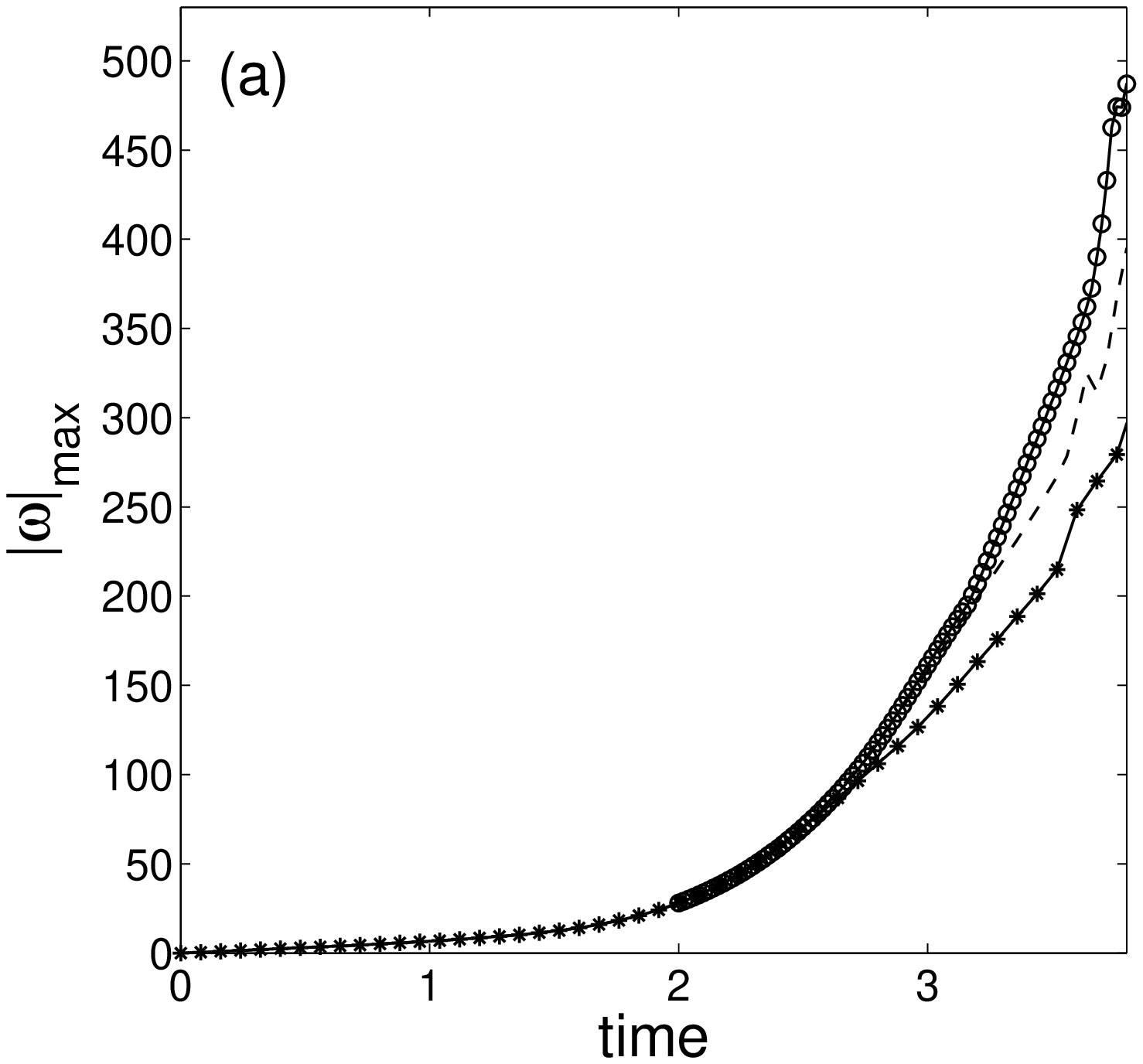}}
\end{minipage}
\begin{minipage}[c]{.489 \linewidth}
\scalebox{1}[1]{\includegraphics[width=\linewidth]{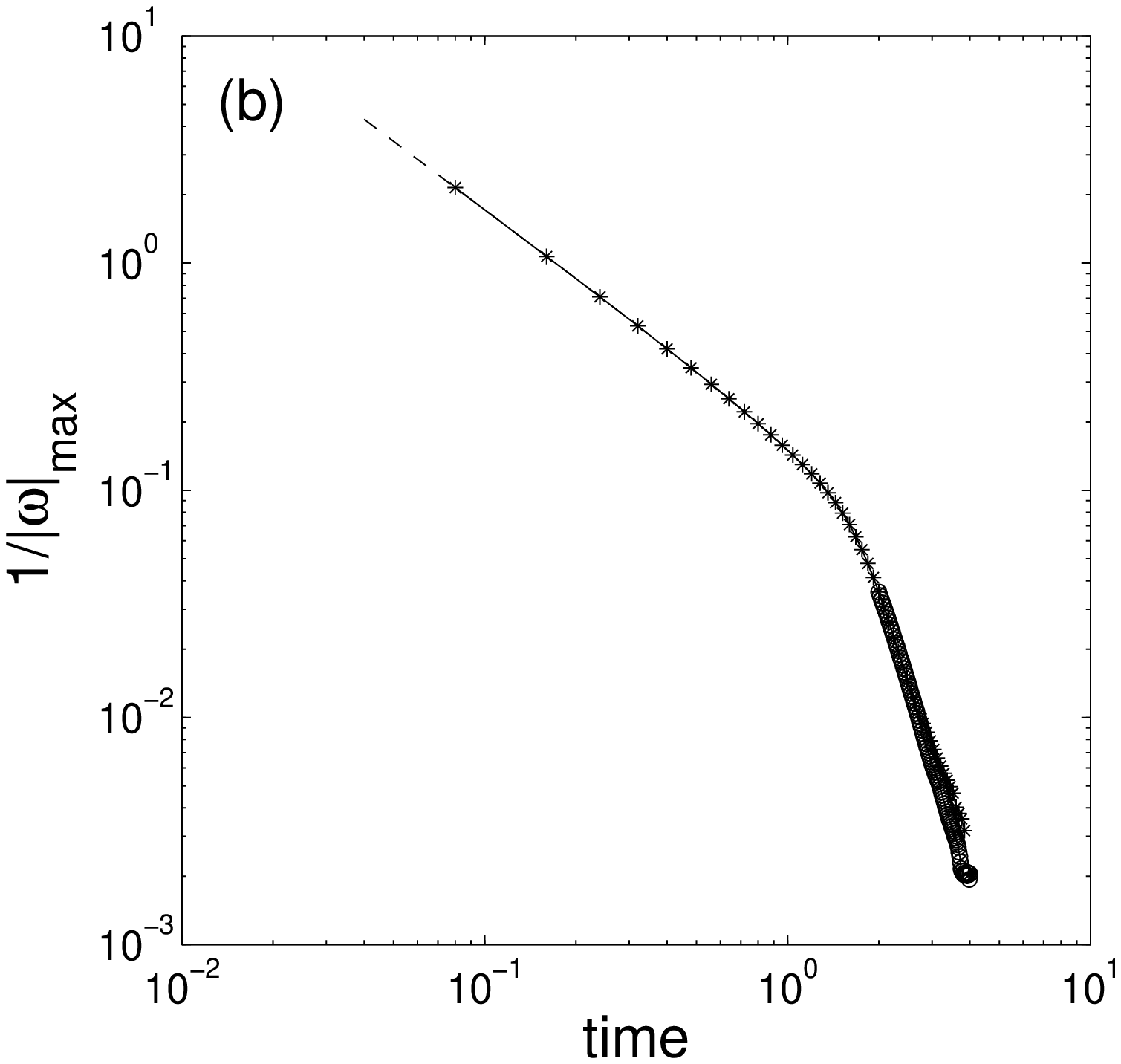}}
\end{minipage}
\begin{minipage}[c]{.489 \linewidth}
\scalebox{1}[1]{\includegraphics[width=\linewidth]{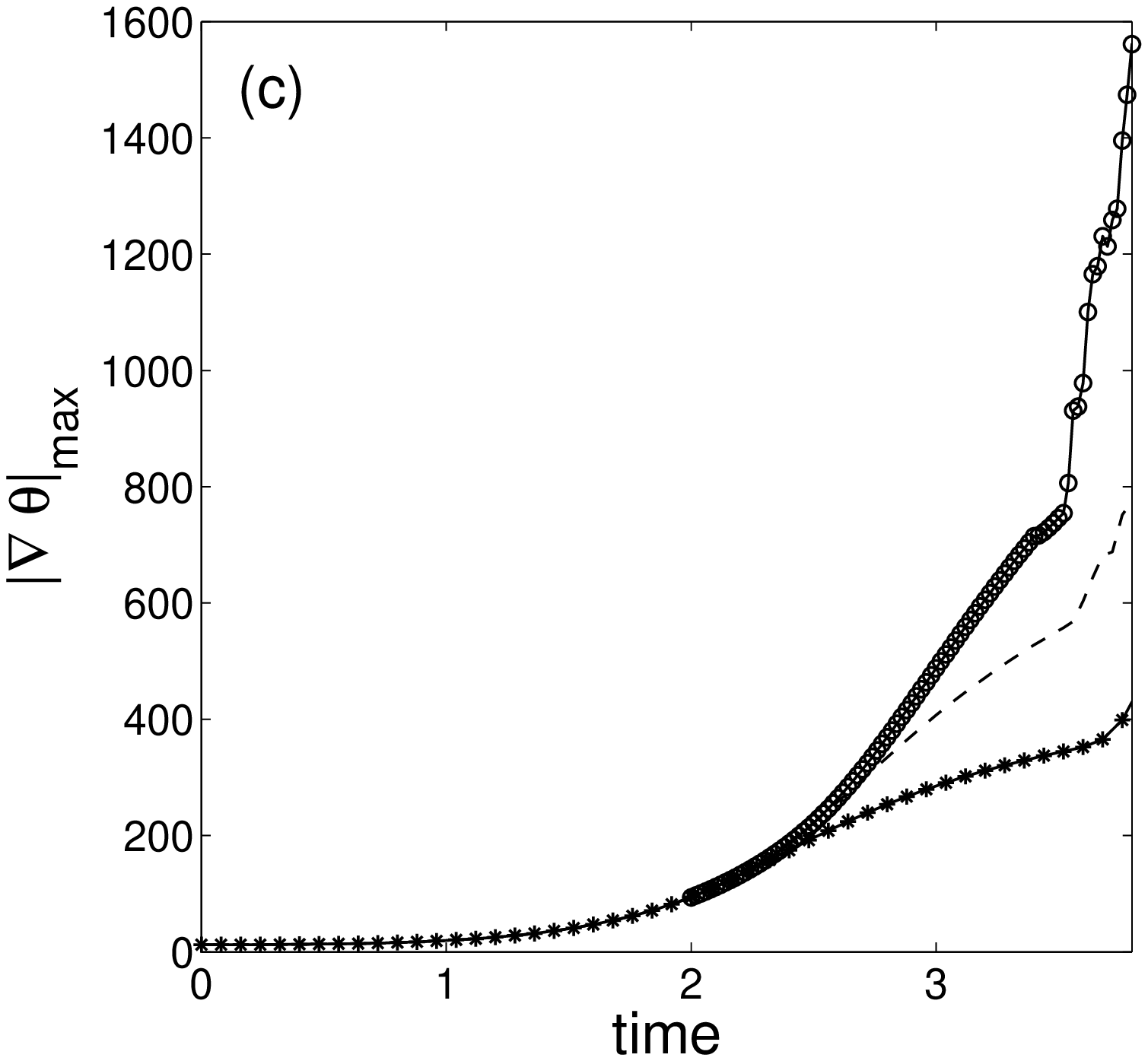}}
\end{minipage}
\begin{minipage}[c]{.489 \linewidth}
\scalebox{1}[1]{\includegraphics[width=\linewidth]{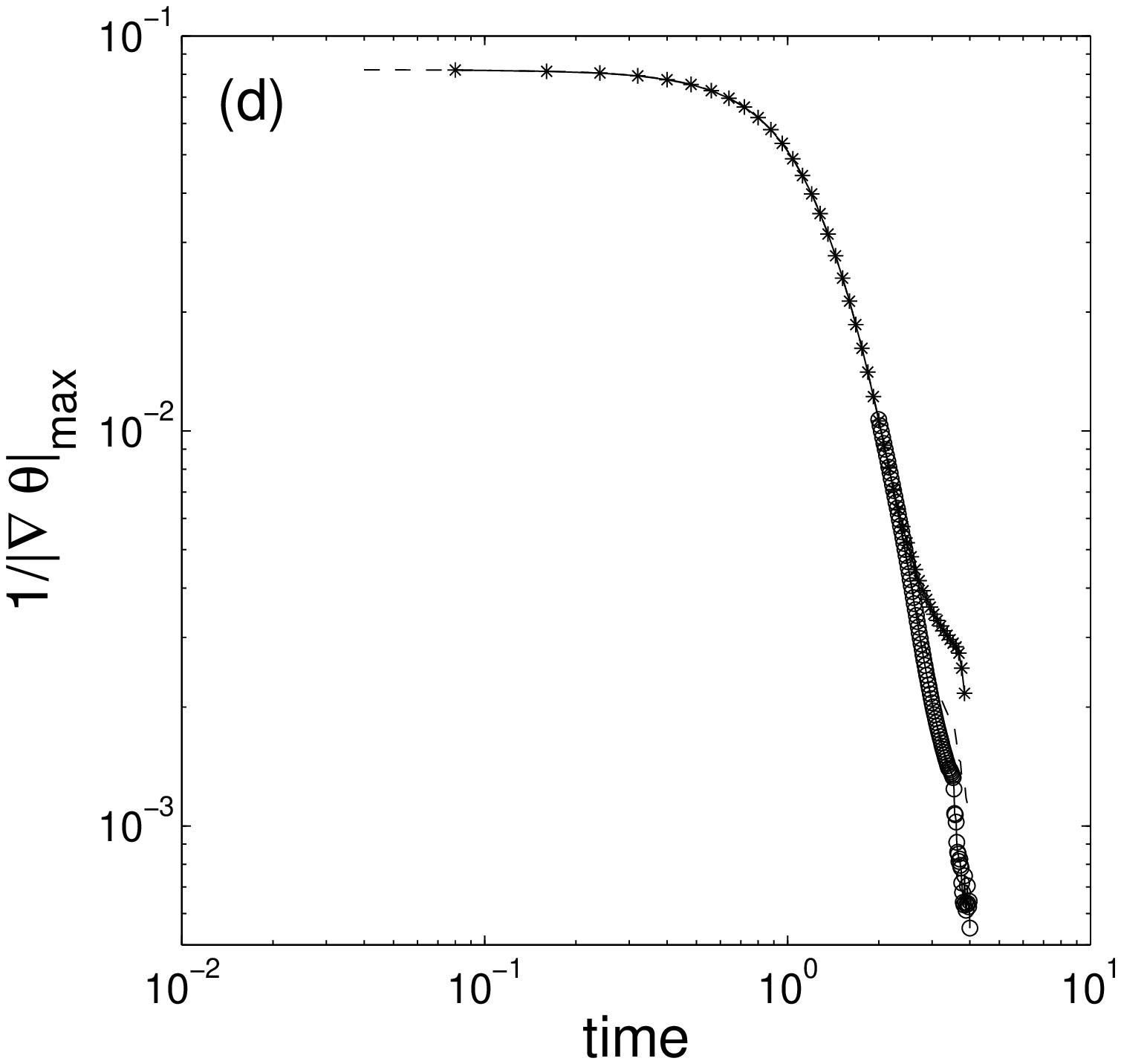}}
\end{minipage}
\end{minipage}
\begin{minipage}[c]{.17 \linewidth}
\caption{\label{fig:Asing} RUN A: Time evolution of $|\omega|_{max}$ and $|\nabla \theta|_{max}$ (solid line: $1024^2$;
dash line: $2048^2$; circle line: $4096^2$).}
\end{minipage}
\end{figure*}

\begin{figure}
\begin{minipage}[c]{.6 \linewidth}
\scalebox{1}[1]{\includegraphics[width=\linewidth]{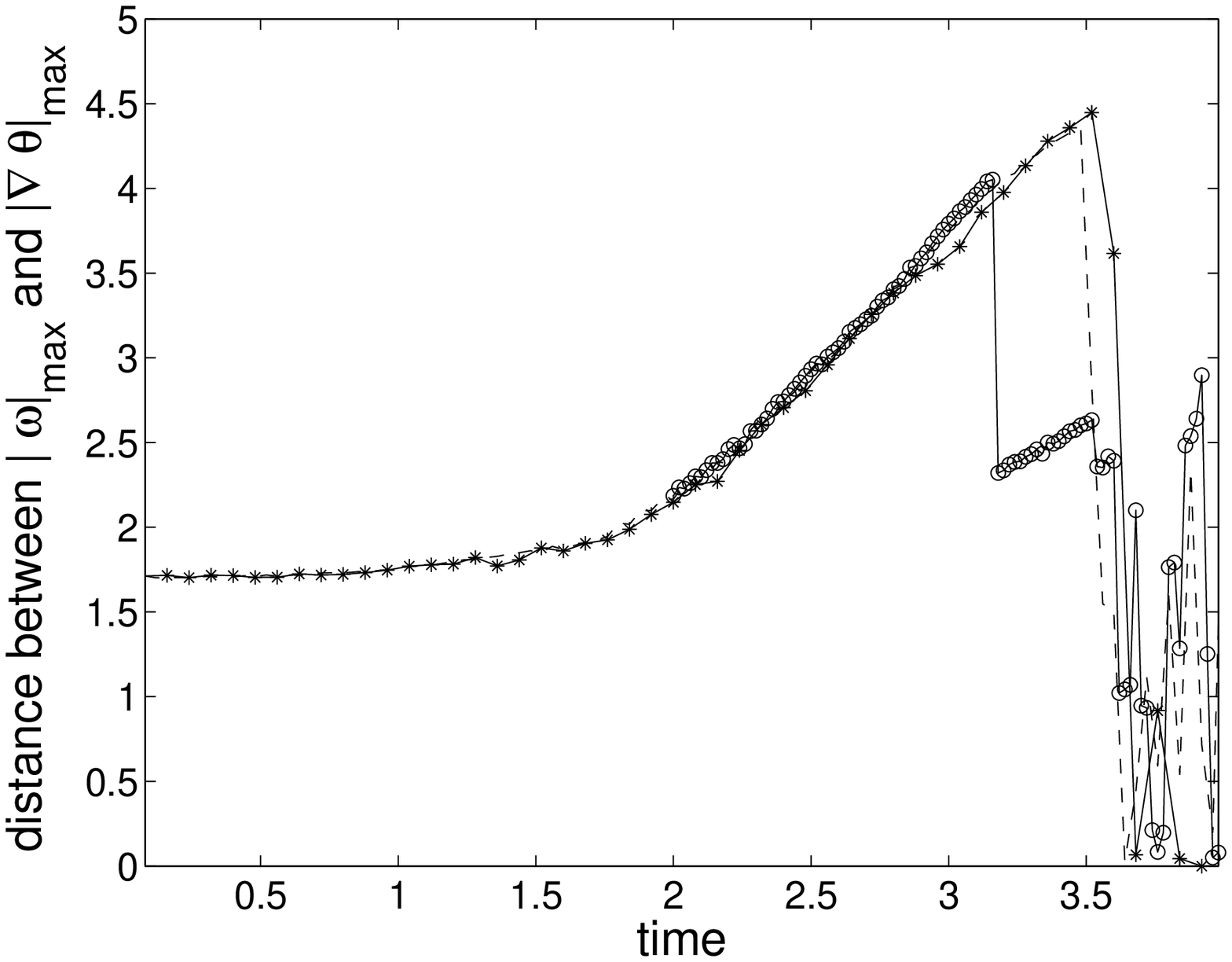}}
\end{minipage}
\caption{\label{fig:Adis} RUN A: The distance between the locations of $|\omega|_{max}$ and $|\nabla \theta|_{max}$ at
different times for three kinds of resolutions (solid line: $1024^2$; dash line: $2048^2$; circle line: $4096^2$).}
\end{figure}

For RUN A, three resolutions are used: $1024^2$, $2048^2$ and $4096^2$.
The corresponding time steps used are 0.00004, 0.00002 and
0.00001 respectively, given by the CFL condition.

At the beginning of the simulations (for all three resolutions),
the cap-like initial temperature field (Fig. \ref{fig:3_initial}(a))
will rise and develop into a ``two-eye'' system.
Driven by the buoyancy, the vorticity field
will also develop from the initially unified zero to a two-eye system, but with one positive and one negative ``eyes''
this time. Figs. \ref{fig:A3dvor} and \ref{fig:A3dp} show the 3D pictures of vorticity and temperature field in the
whole domain, and it is clear that the peak values of vorticity are located along the edge of the ``eyes.''

The system is basically symmetric respect to $x = \pi$ if round-off
error does not play a role, so there will have two
$|\omega|_{max}$ locations: one positive and one negative. It is
worth mentioning that round-off error will spoil this
symmetry when very high resolutions are adopted
(see the discussion in Appendix). Our highest resolution for RUN A is
$4096^2$, and the symmetry is well maintained before $t = 3.6$. After $t = 3.6$, the filter spoils the
singularity-forming mechanism, and the values of $|\omega|_{max}$ begin to drop when $t$ becomes larger. In the
following, we will only use the positive half of the vorticity field (i.e. $x \in [\pi, 2\pi]$) for the divergence
analysis.

The ``+'' symbol in Figs. \ref{fig:Azoomv} indicates
the locations of $|\omega|_{max}$ at different times. The ``+'' is
around $y = 3$ in the beginning, and suddenly jumps to $y \approx 4.5$ after $t = 3.1$. The flow field has no
significant change around $t = 3.1$, and the front of the bubble
rises continuously as what happens before that
time.

The ``*'' symbol in Figs. \ref{fig:Azoomp} indicates the
locations of $|\nabla \theta|_{max}$ at different times. It
should be noticed that the tail left behind the rising front is forming an ``eye'' after $t = 2.5$, and there is a
smooth filament connecting the ``eye'' and the head of the bubble.

The ``$*$'' symbol jumps to the edges of the forming ``eye''
around $t \approx 3.6$, roughly at the same time when the
original smooth filament breaks up into many even smaller ``eyes'' (see Figs. \ref{fig:Around}). This is because our
simulation begins to become under-resolved.
The filter begins to dramatically reduce the peak values of those
$\delta$-like functions. The ``*'' symbols can not represent the exact maximum locations of the 2D Boussinesq solution
anymore. The edge of temperature contour is very
smooth before the collapsing time, and the front part of the rising
bubble and the two eyes left behind (Figs. \ref{fig:A3dp}) stretch the filament connecting them.

Questions arise as to how effectively the large local quantity (in our case, the vorticity and temperature gradients)
can be resolved and how much the results vary with resolution. It is evident that for the extreme hypothetical case,
when one of the local quantities is a delta function, it can only be resolved with infinite resolution. To resolve an ideal shock
wave (with zero thickness), for example, in the absence of any
smoothing, needs infinite number of grid points.
In the following, we will check
several other aspects on the effectiveness of
the numerical  simulations, mainly by
finding some quantity properties useful in demonstrating the effect of the
mesh refinement, or identifying the trend
when high resolutions are adopted.

First, we check the following three values which are time independent
due to the divergence-free constraint, the
doubly-periodic condition and the inviscid transport equation
(Eq. (\ref{boussinesq-1})):
\begin{itemize}
\item
$T_2(t) = \int^{2\pi}_{0}\int^{2\pi}_{0} \theta^2(x,y,t) dxdy$,
\item
$T_4(t) = \int^{2\pi}_{0}\int^{2\pi}_{0} \theta^4(x,y,t) dxdy$, and
\item
$\theta_{max}(t)$
\end{itemize}
Fig. \ref{fig:Aerror} shows that the global average quantities
are well conserved within $1\%$ error for
$4096^2$ and $2048^2$ resolutions throughout the entire
simulation period $t \in [0,4]$. To save the computation time, our
$4096^2$ run starts from $t = 2.0$ using the intermediate result of
$2048^2$, so for $4096^2$ run, those errors are
defined as ${(T_2(2.0)-T_2(t))}/{T_2(2.0)}$,
${(T_4(2.0)-T_4(t))}/{T_4(2.0)}$ and $|\theta_{max}(2.0) -
\theta_{max}(t)|/|\theta_{max}(2.0)|$.

The $1024^2$ run has a very poor performance after $t = 3.8$ because none of
these relative errors are below 10\%. The performance of $2048^2$
run is almost as good as on $4096^2$ grid by considering $T_2$ and $T_4$,
with under 1\% relative errors. However, the relative errors on
$\theta_{max}$ of $4096^2$ are always below $10^{-4}$,
which are only about 2\% of the errors in $2048^2$ simulation
after $t = 3.5$ (Figs. \ref{fig:Aerror}(c)).
These errors give some indicators on
how far the numerical solution is away from the real one.

Figs. \ref{fig:Aerror}(a) and (b) show that $T_2$ and $T_4$
errors of $4096^2$ run are always lower than those given by the $1024^2$
and $2048^2$ simulations. However, due to the round-off error
in our calculations (see Appendix), this is not the case
for $\theta_{max}$ error (Figs. \ref{fig:Aerror}(c)),
because the errors associated with the  $4096^2$
resolution seems to be at the same level of
those with the  $2048^2$ resolution
from $t=2.0$ to $t=2.6$. When $t$ is close to the blow-up time,
the $4096^2$ run conserves the value
of $\theta_{max}$ much better than other runs.
In fact, when $t\to T_c$, the truncation error will overwhelm the
influence of machine precision, and higher resolutions
have more advantages than lower ones. For future simulations,
however, we may have to take some measures to control the round-off
error when refined grids are used, say,
the quadruple precision (the machine
accuracy $\epsilon = 10^{-32}$) may be used to replace the present
double-precision ($\epsilon = 10^{-16}$). On the other hand,
the present results of $2048^2$ and $4096^2$ simulations
seem to be accurate enough to draw some conclusions in the blow-up
analysis.

Figs. \ref{fig:Asing}(a) and (c) show the time evolutions for $|\omega|_{max}$ and $|\nabla \theta|_{max}$ with
different resolutions. We re-plot these maximum value evolutions in Figs. \ref{fig:Asing}(b) and (d) with logarithmic
scales. It seems that the growth of $|\omega|_{max}$ and $|\nabla \theta|_{max}$ in the $4096^2$ run terminates in a
finite time with $|\omega|_{max} \sim {(T_c-t)}^{-1.23}$ and crudely $|\nabla \theta|_{max} \sim {(T_c - t)}^{-2.48}$.
The corresponding predicted blow-up time is $T_c = 3.72$. Even for $2048^2$ run, there is an obvious trend towards the
singularity.

However, there are still two facts that cause suspects of this conclusion:
\begin{itemize}
\item The simulation never reaches the blow up time $T_c$;
\item Even for the $4096^2$ run, we can not say that we have resolved
the problem confidently because the time-evolution
curves for $|\omega|_{max}$ and $|\nabla \theta|_{max}$ show quite large
differences between high and low resolutions at
late stages of the simulations.
\end{itemize}
The first point is unavoidable due to the discrete nature of
the numerical simulations. The second point may be improved by employing
higher resolution, or, at least we can extend the overlapping
region to a time closer to $T_c$.
For example, the $|\omega|_{max}(t)$ lines of $1024^2$ and
$2048^2$ runs diverge around $t = 2.7$, but the $2048^2$ and
$4096^2$ curves diverge around $t = 3.2$. However, the effort
in this direction may be useless because, as explained in the
following paragraph, we can not get a clearer physical picture
of the singularity forming mechanism even we use higher
resolution.

Physically speaking, if a singularity is about to form
at $t = T_c$ on the point $(x_c,y_c)$, the locations of
$|\omega|_{max}$ and $|\nabla \theta|_{max}$ should approach
$(x_c,y_c)$ when $t \to T_c$. This means the distance
between the locations of these two peak values should
become smaller and smaller as $t \to  T_c$. Although it is
impossible for any numerical solver to reach $T_c$,
we can use this property to check whether a numerical
simulation reflects the physical mechanism properly.
Fig. \ref{fig:Adis} shows that the locations of $|\omega|_{max}$
and $|\nabla \theta|_{max}$ in RUN A initially stay far
away from each other (about 1.7 length unit in the $[0, 2 \pi]
\times [0, 2 \pi]$ domain) and they do not get closer when
the ``singular'' time is approaching \footnote{The location of
$|\nabla \theta|_{max}$ will cross the boundary at $t \approx 3.2$,
so we measure the distance on the $[0, 2 \pi]
\times [1.5, 1.5 + 2\pi]$ domain instead of on $[0, 2 \pi] \times [0,2\pi]$.}.
The distance are noticeably increased
after $t \approx 1.7$ for all three resolutions.
When  $t \in [3.2, 3.5]$, it is even wider (see also the specific
locations of those maximum values in
Figs. \ref{fig:Azoomv} and \ref{fig:Azoomp}).
The physical mechanism behind it is
very unclear, maybe because the numerical code is trying to
preserve the symmetry from the very beginning of flow
pattern (Chapter 2 of ~\cite{fri95}),
accompanied by the singularity forming mechanism.
It seems that there are at least
two locations at which the singularity is formed
simultaneously at $x \in [\pi, 2\pi]$, fighting for the locations of global
maximum values. Because the divergent conclusion drawn from
Figs. \ref{fig:Asing}(b) and (d) means nothing unless higher resolutions
are adopted, we should consider some other initial condition which may
reveal more physical phenomena more effectively.
The results of RUN A seem too complicated to
be analyzed.

To sum up for this subsection, we basically end up with
similar conclusion as in \cite{e94} in the sense that the edge of
the ``cap'' is more dangerous than the front.
Although our code can resolve the 2D Boussinesq equations accurately
until about $t \approx 3.5$ on the $4096^2$ grid, we can not
obtain a clear physical picture relevant to the singularity issue for the
3D Euler equations. RUN A can only be used to reveal some
singular phenomena in the 2D Boussinesq equations.

\subsection{Numerical results for RUN C}

In RUN C, we use four sets of grids, namely
$1024^2$, $2048^2$, $4096^2$ and $6144^2$. The corresponding
 time steps are 0.00004, 0.00002,
0.00001 and 0.000008 respectively (for $4096^2$ run, we use $\Delta t = 0.00002$ for $t < 0.22$, and 0.00001 for $t
\geq 0.22$ to save the computation time).

\begin{figure*}
\begin{minipage}[c]{.3 \linewidth}
\scalebox{1}[1.2]{\includegraphics[width=\linewidth]{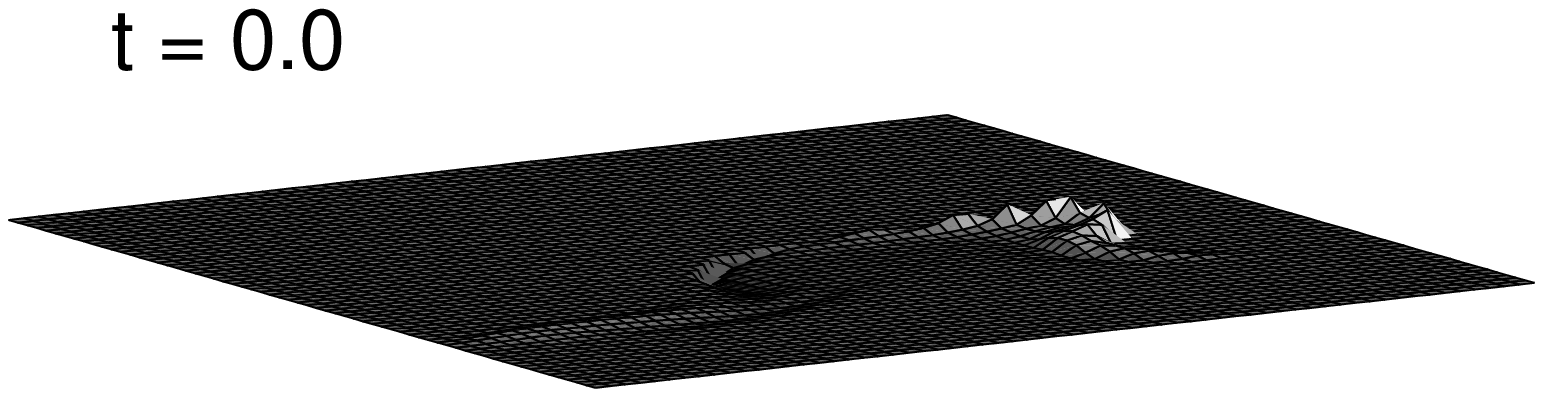}}
\end{minipage}
\begin{minipage}[c]{.3 \linewidth}
\scalebox{1}[1.2]{\includegraphics[width=\linewidth]{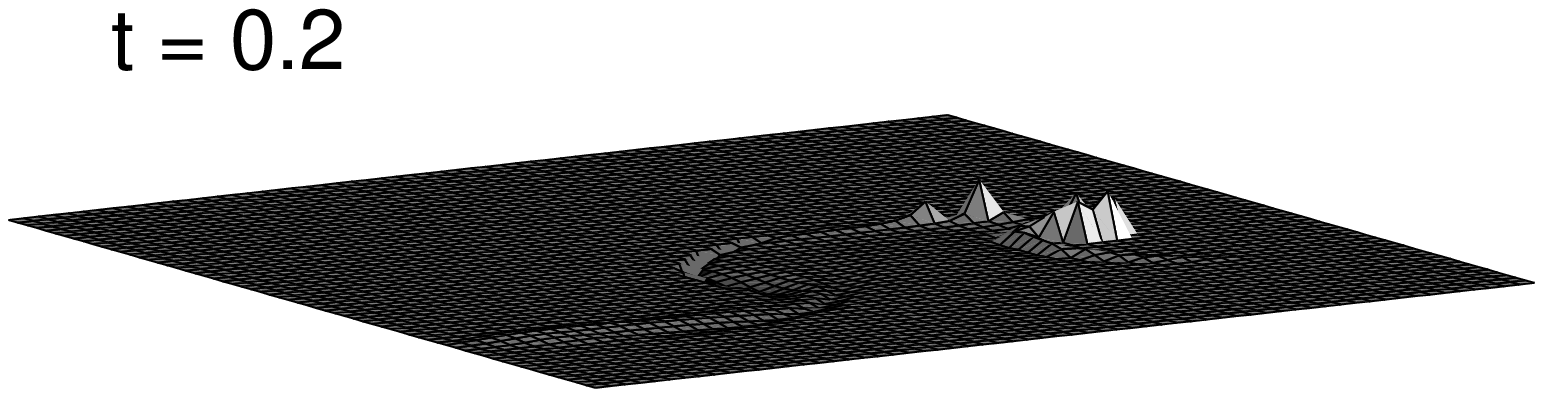}}
\end{minipage}
\begin{minipage}[c]{.3 \linewidth}
\scalebox{1}[1.2]{\includegraphics[width=\linewidth]{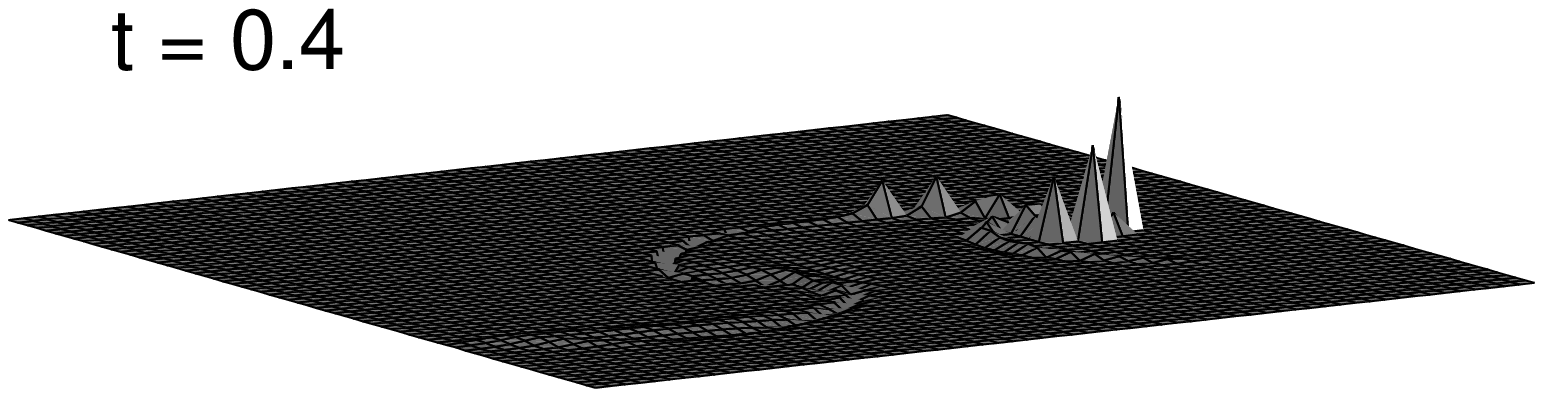}}
\end{minipage}
\begin{minipage}[c]{.3 \linewidth}
\scalebox{1}[1.2]{\includegraphics[width=\linewidth]{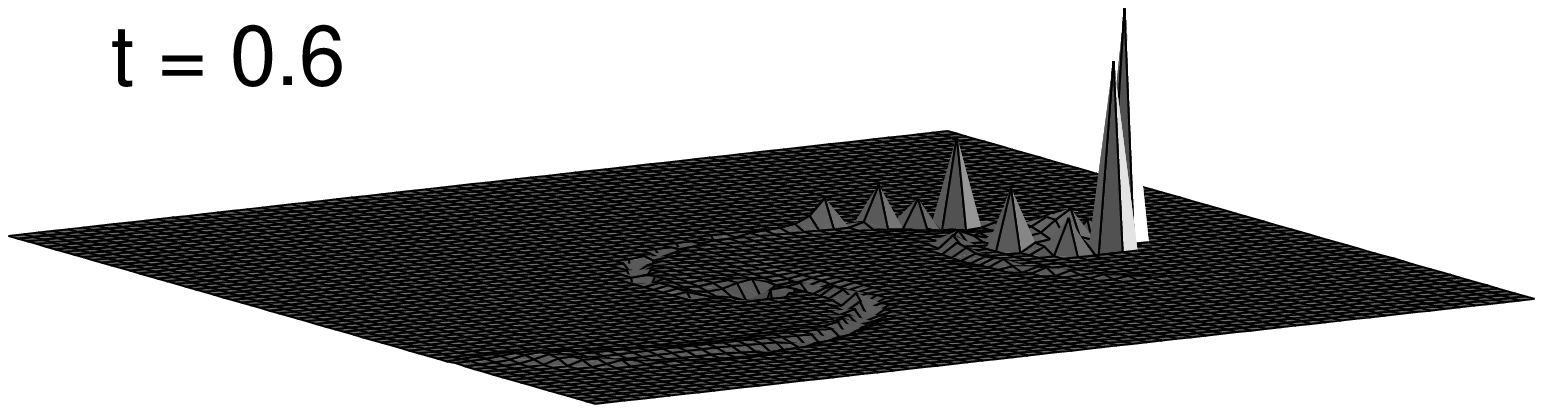}}
\end{minipage}
\begin{minipage}[c]{.3 \linewidth}
\scalebox{1}[1.2]{\includegraphics[width=\linewidth]{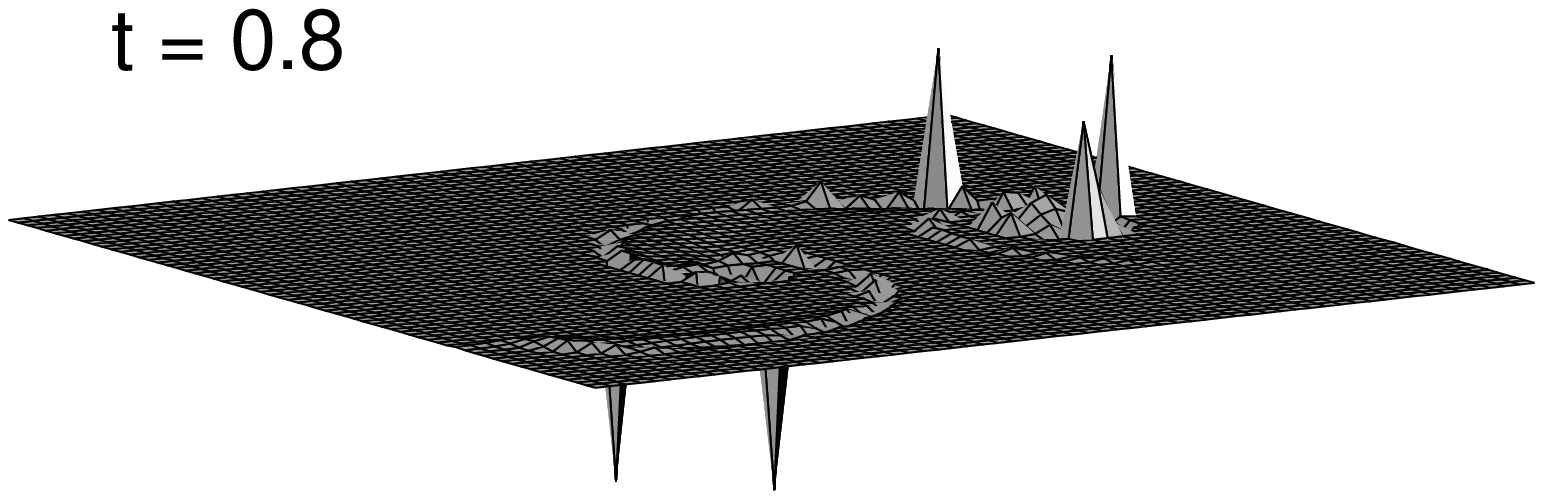}}
\end{minipage}
\begin{minipage}[c]{.3 \linewidth}
\scalebox{1}[1.2]{\includegraphics[width=\linewidth]{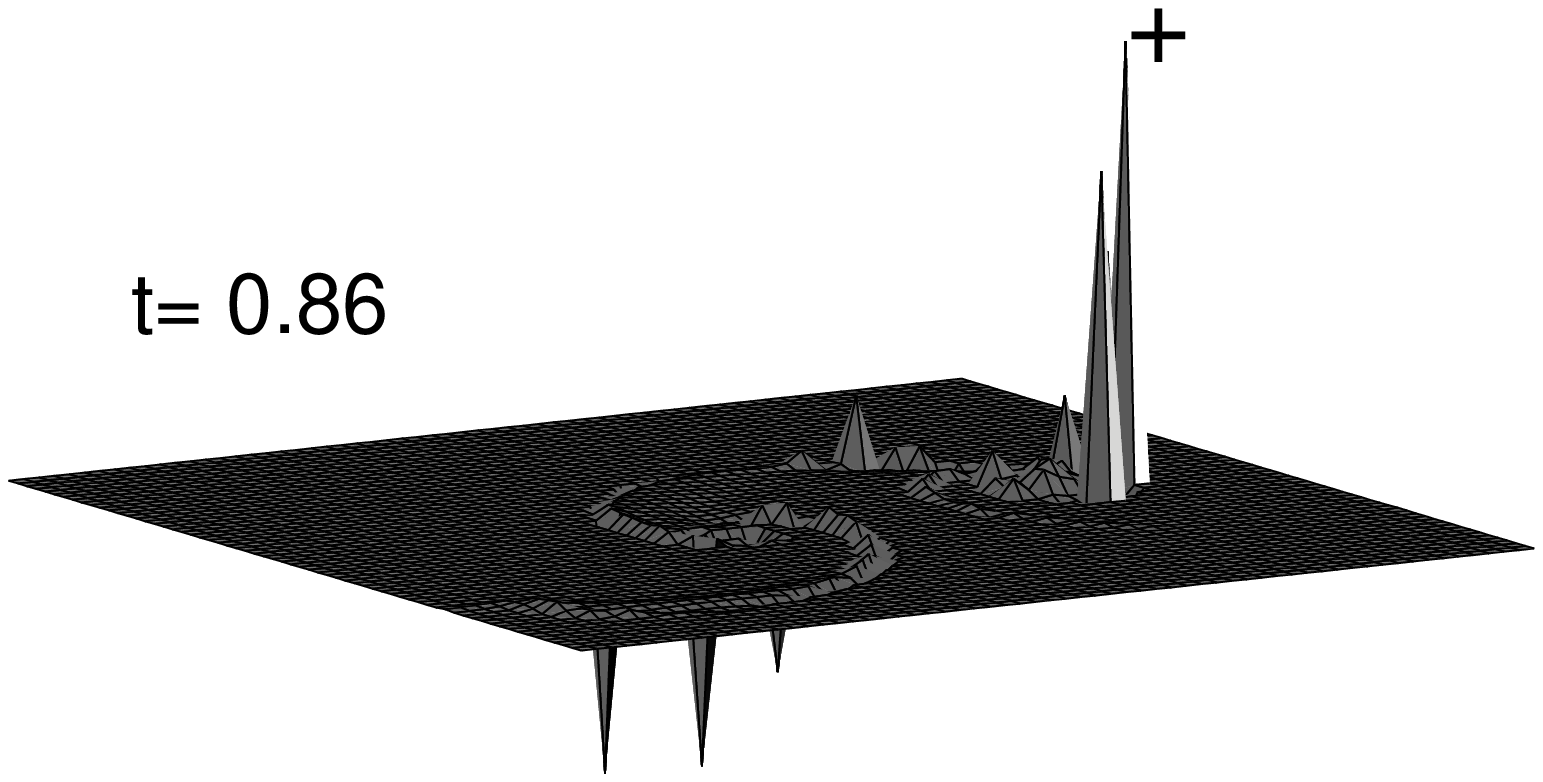}}
\end{minipage}
\caption{\label{fig:C3dvor} The 3D perspective plots of vorticity for RUN C with the $6144^2$ grid.}
\end{figure*}
\begin{figure*}
\begin{minipage}[c]{.99 \linewidth}
\scalebox{1}[1]{\includegraphics[width=\linewidth]{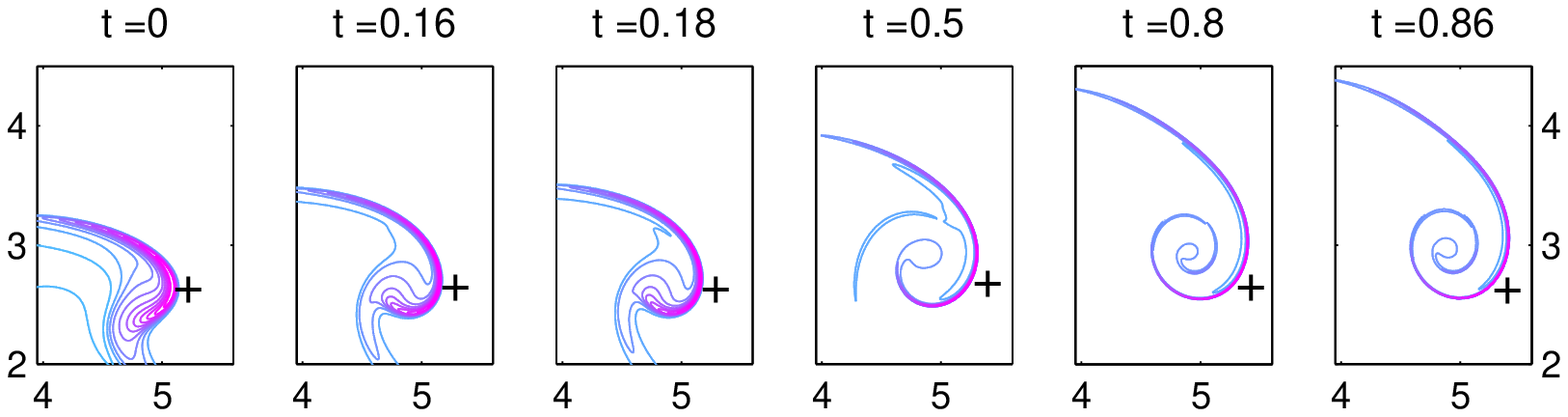}}
\end{minipage}
\begin{minipage}[c]{.99 \linewidth}
\scalebox{1}[1]{\includegraphics[width=\linewidth]{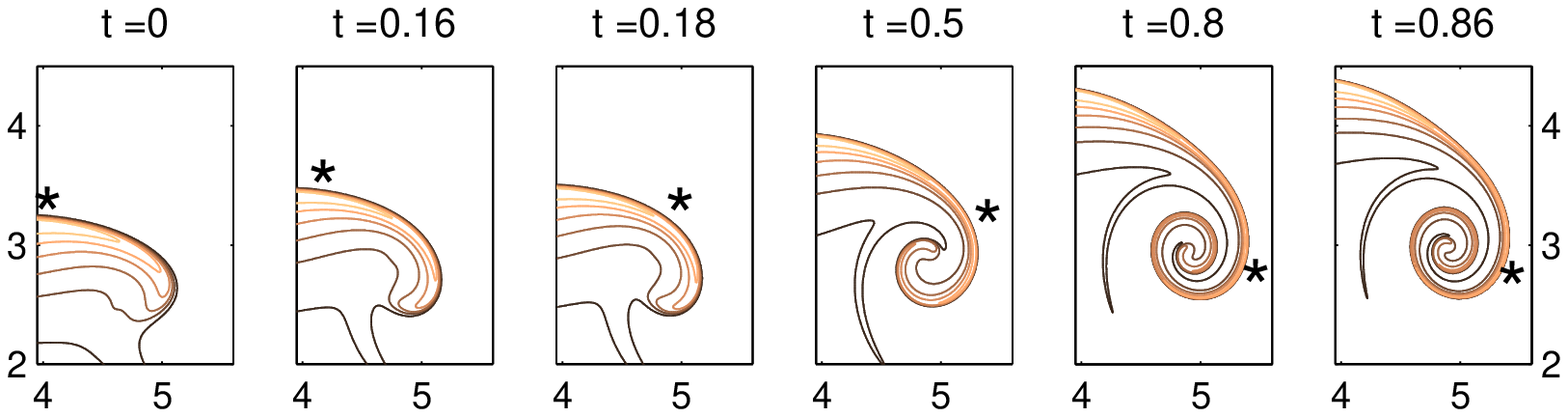}}
\end{minipage}
\caption{\label{fig:Czoom} (Color online). Zoom-in contour plots of vorticity (the first row) and temperature (the
second row) in RUN C at different times with the resolution of $4096^2$. The ``+'' and ``*'' indicate the location of
$|\omega|_{max}$ and $|\nabla \theta|_{max}|$ in the whole $[0, 2 \pi] \times [0, 2 \pi]$ domain, respectively.}
\end{figure*}
\begin{figure*}
\begin{minipage}[c]{.49 \linewidth}
\scalebox{1}[1.2]{\includegraphics[width=\linewidth]{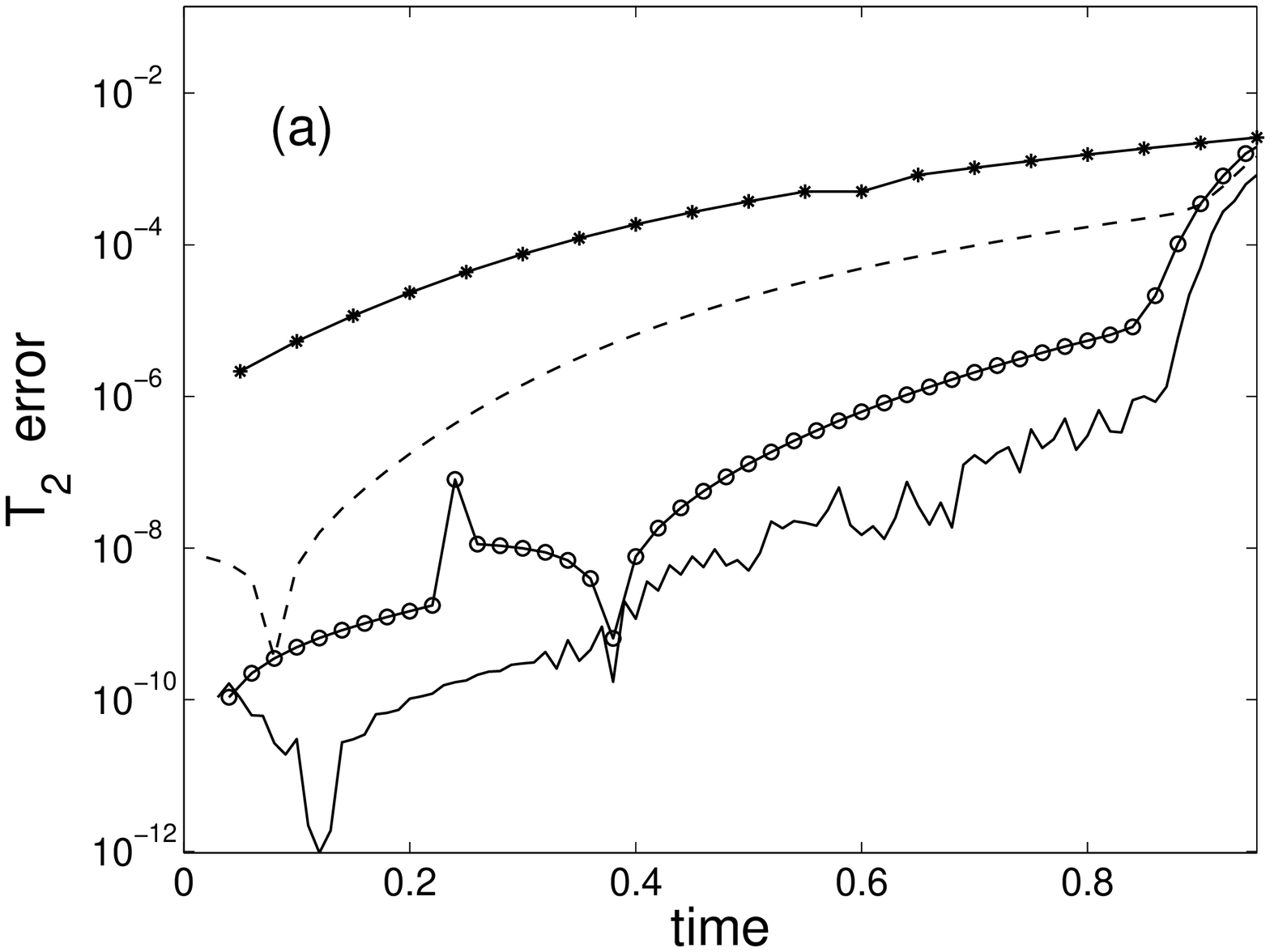}}
\end{minipage}
\begin{minipage}[c]{.49 \linewidth}
\scalebox{1}[1.2]{\includegraphics[width=\linewidth]{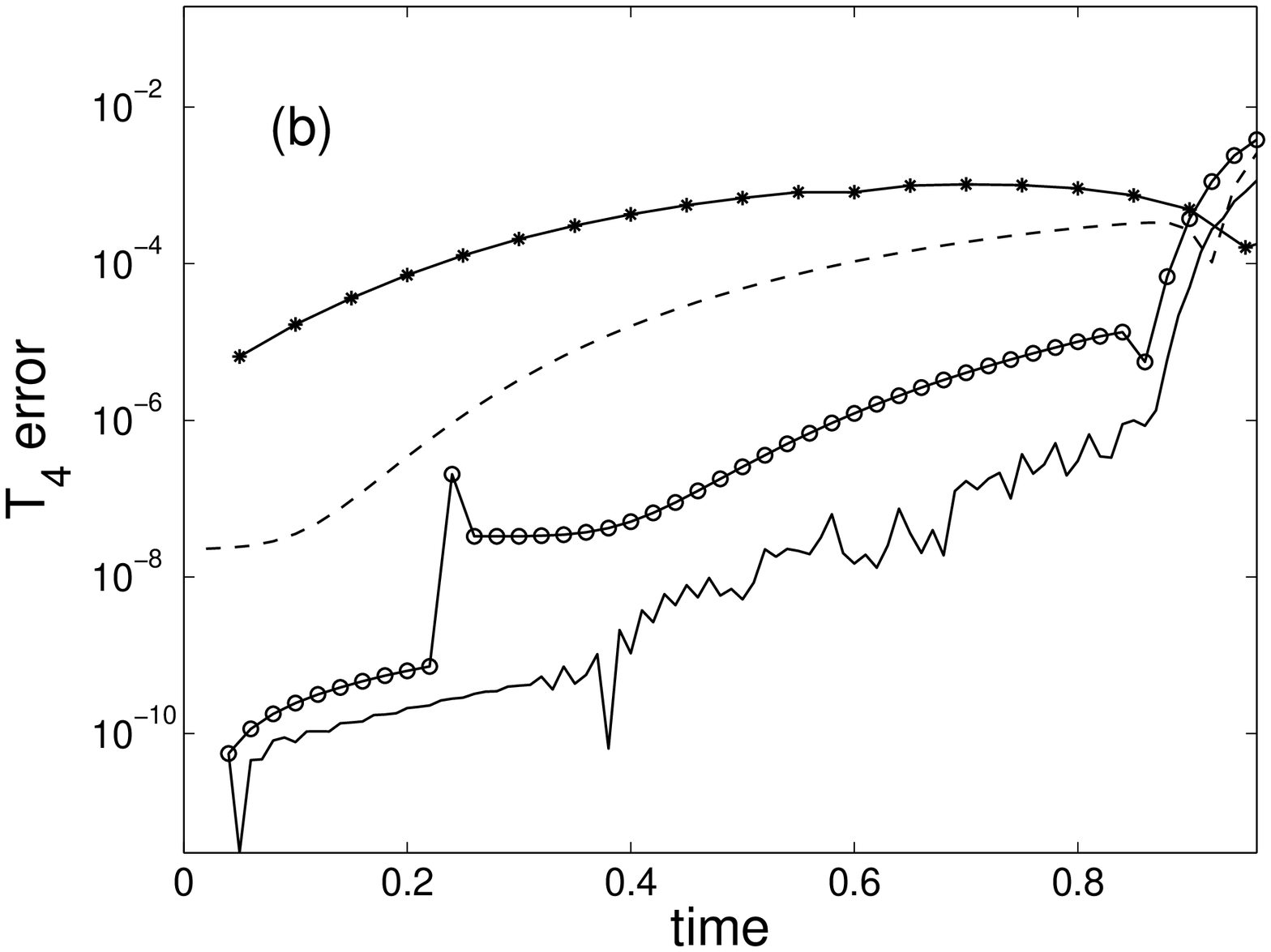}}
\end{minipage}
\begin{minipage}[c]{.49 \linewidth}
\scalebox{1}[1.2]{\includegraphics[width=\linewidth]{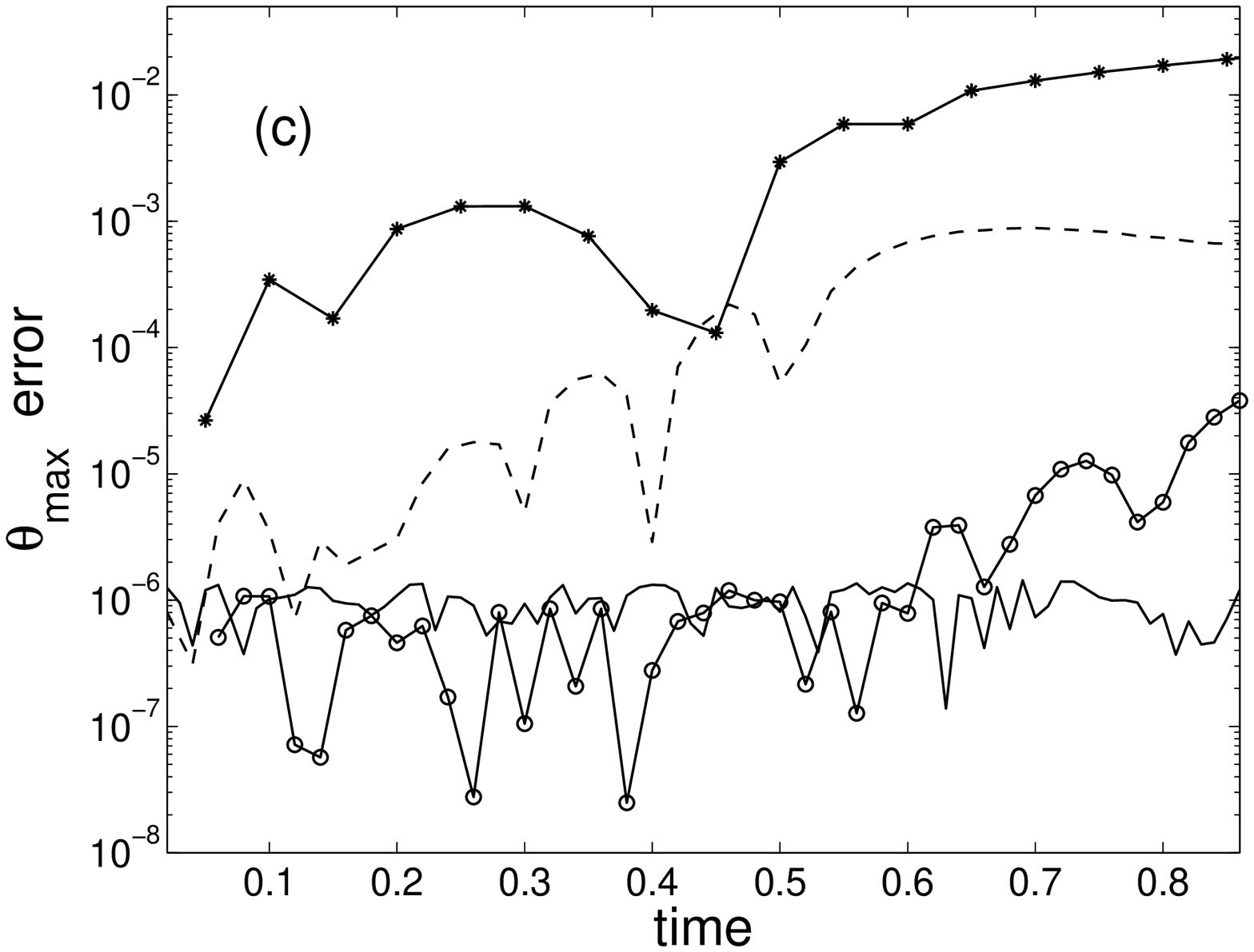}}
\end{minipage}
\begin{minipage}[c]{.49 \linewidth}
\begin{minipage}[c]{.98 \linewidth}
\caption{\label{fig:Cerror} RUN C: Time evolutions of the relative errors for $T_2$, $T_4$ and $\theta_{max}$ (star
line: $1024^2$; dash line: $2048^2$; circle line: $4096^2$; solid line: $6144^2$).}
\end{minipage}
\end{minipage}
\end{figure*}

\begin{figure}
\begin{minipage}[c]{0.65 \linewidth}
\scalebox{1}[1.1]{\includegraphics[width=\linewidth]{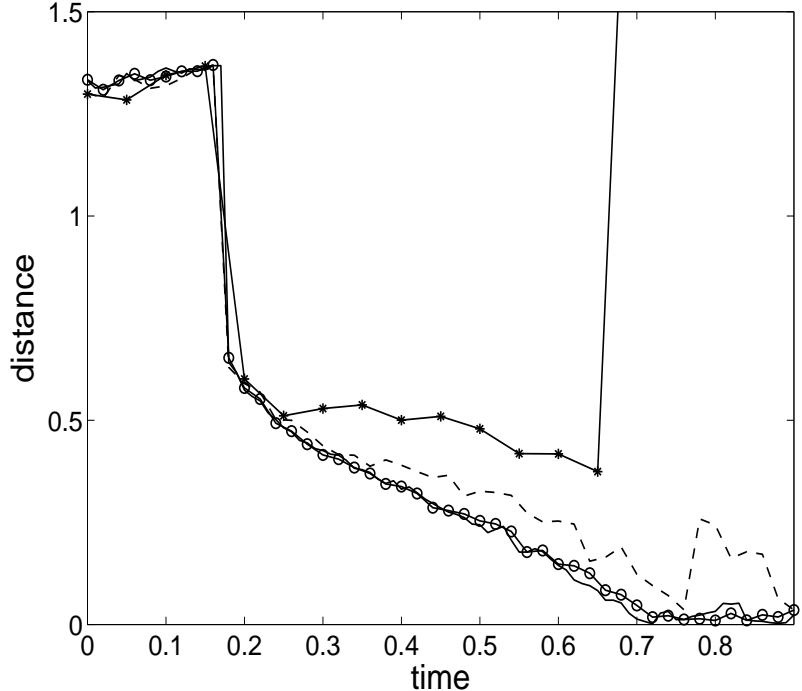}}
\end{minipage}
\caption{\label{fig:Cdis}RUN C: The distance between the locations of $|\omega|_{max}$ and $|\nabla \theta|_{max}$ at
different times with four resolutions (star line: $1024^2$; dash line: $2048^2$; circle line: $4096^2$; solid line:
$6144^2$).}
\end{figure}

\begin{figure*}
\begin{minipage}[c]{.82 \linewidth}
\begin{minipage}[c]{.49 \linewidth}
\scalebox{1}[1]{\includegraphics[width=\linewidth]{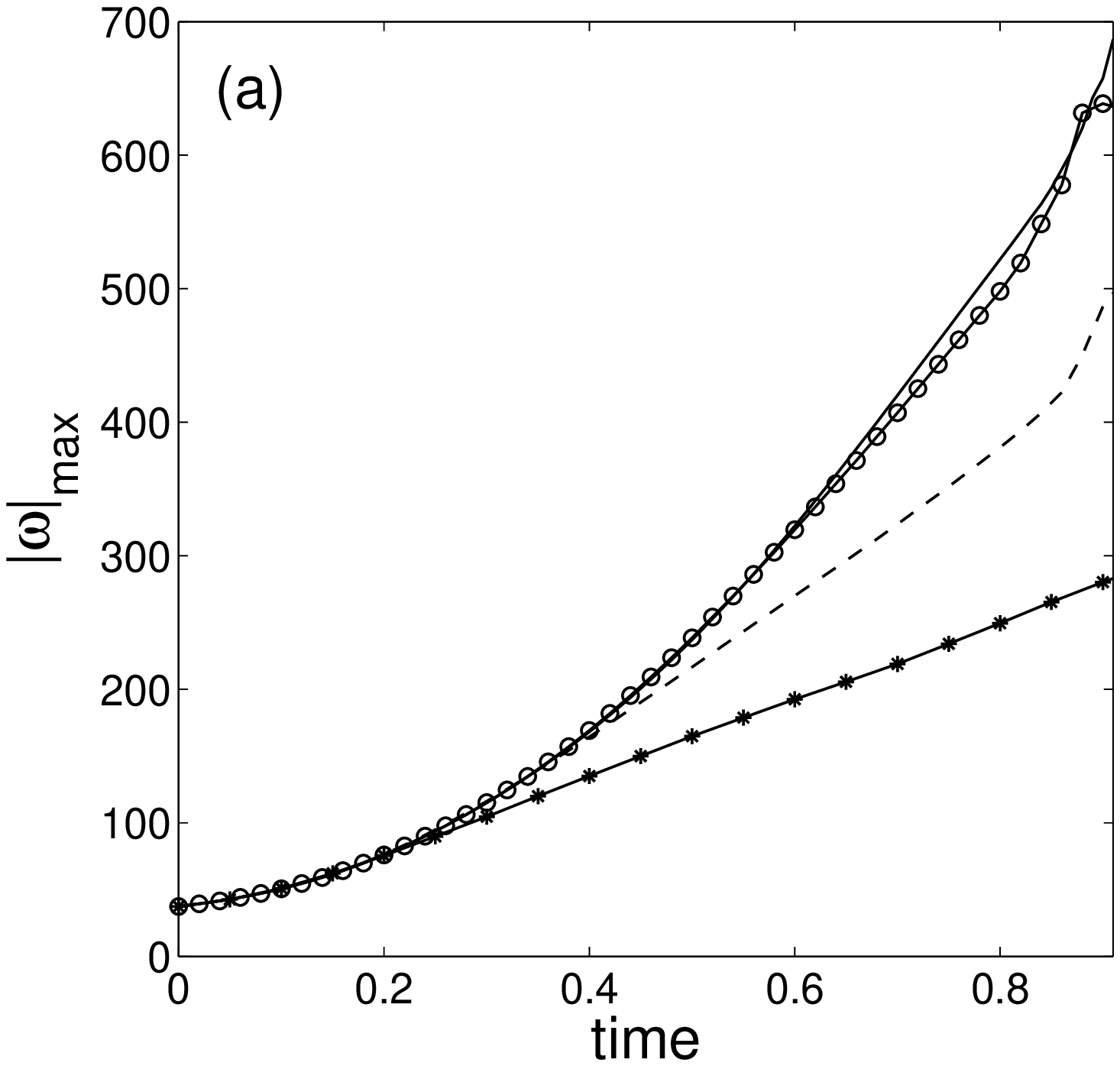}}
\end{minipage}
\begin{minipage}[c]{.49 \linewidth}
\scalebox{1}[1]{\includegraphics[width=\linewidth]{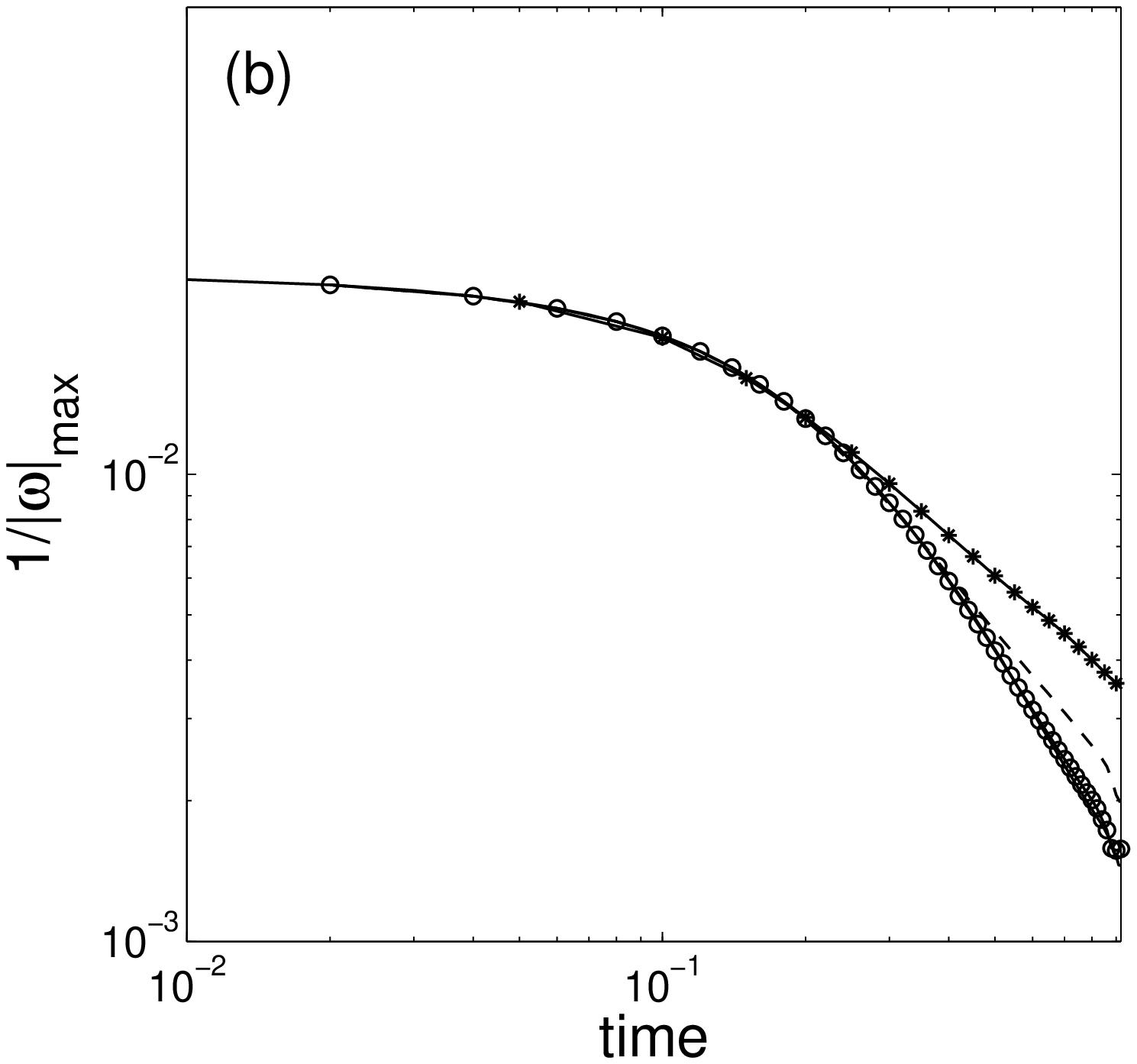}}
\end{minipage}
\begin{minipage}[c]{.49 \linewidth}
\scalebox{1}[1]{\includegraphics[width=\linewidth]{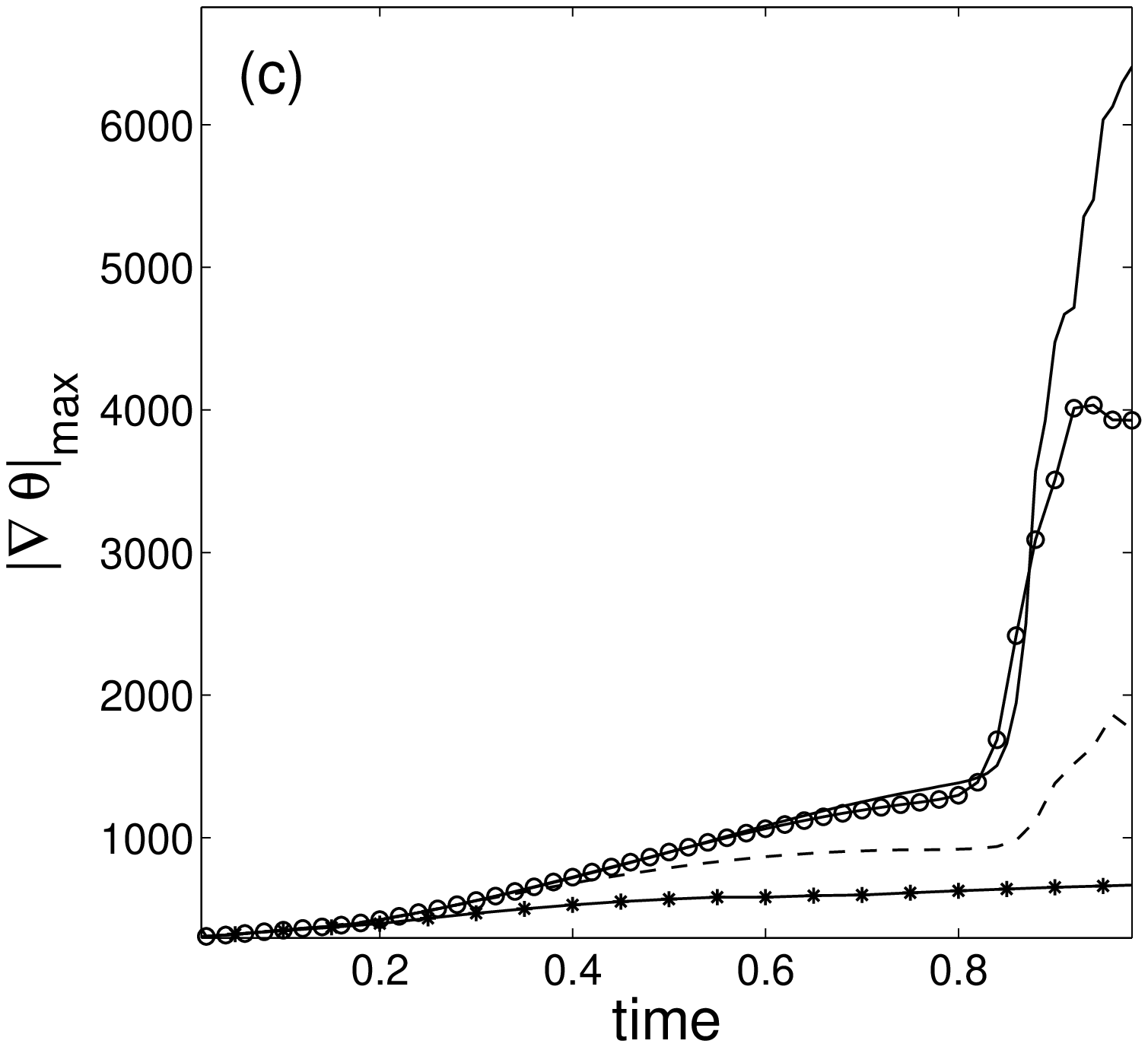}}
\end{minipage}
\begin{minipage}[c]{.49 \linewidth}
\scalebox{1}[1]{\includegraphics[width=\linewidth]{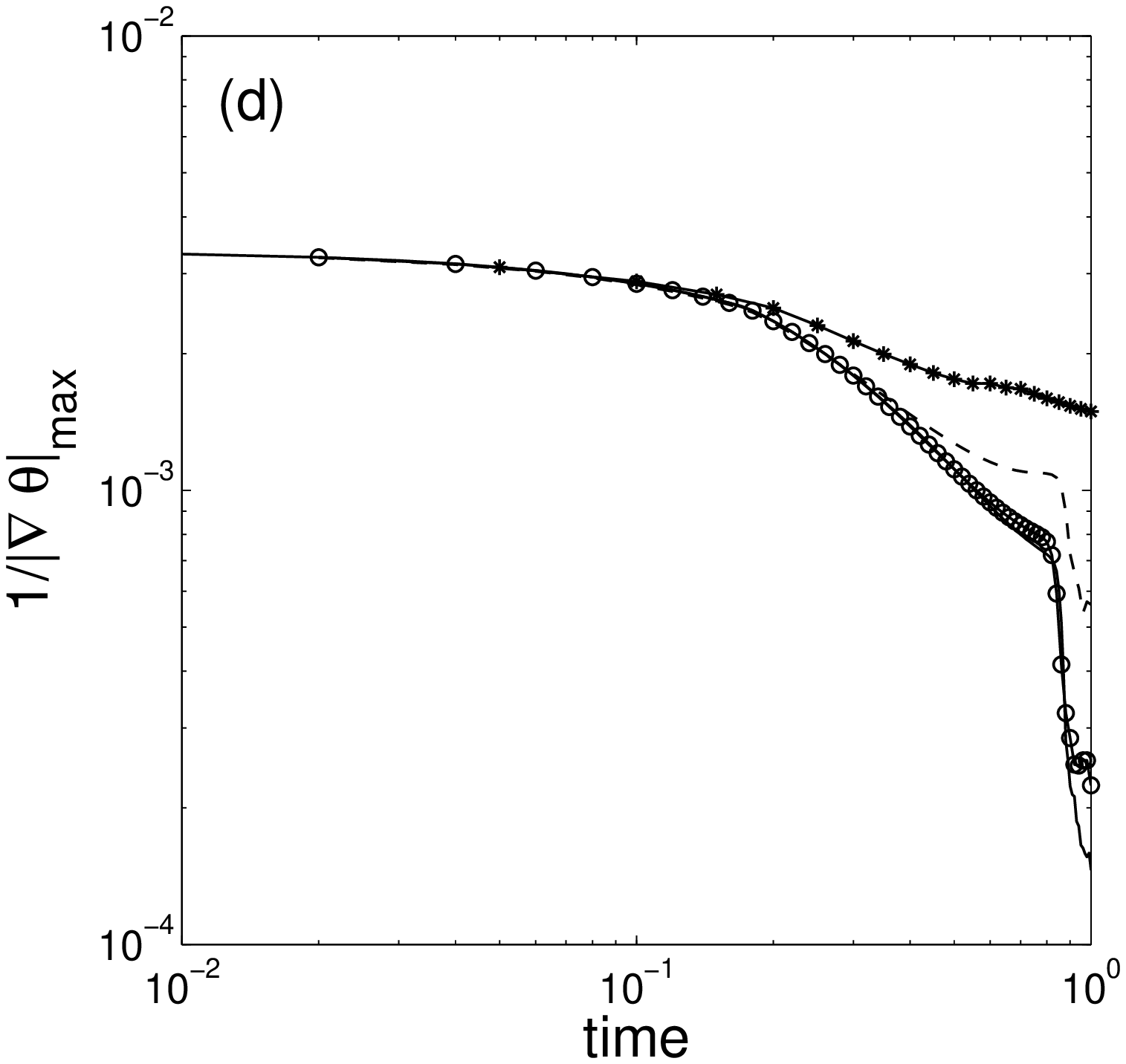}}
\end{minipage}
\end{minipage}
\begin{minipage}[c]{.17 \linewidth}
\caption{\label{fig:Csing} RUN C: Time evolution of $|\omega|_{max}$ and $|\nabla \theta|_{max}$ (star line: $1024^2$;
dash line: $2048^2$; circle line: $4096^2$; solid line: $6144^2$). }
\end{minipage}
\end{figure*}

Fig. \ref{fig:C3dvor}  shows the 3D pictures of vorticity in the whole domain of RUN C. The rising bubble also develops
into two ``eyes'' like that in RUN A. In RUN C, as a whole,
the positive ``eye'' has much larger absolute values for $|\omega|$
and $|\nabla \theta|$ than those of the negative ``eye.''
There is no symmetry with respect to $x = \pi$ anymore. In the
following, we will only exam the details around the positive
``eye'' with a particular attention to the locations of the global maximal
$|\omega|$ and $|\nabla \theta|$ (see Fig. \ref{fig:Czoom}).

A noticeable difference between Fig. \ref{fig:Czoom} and Fig.
\ref{fig:Azoomv} is about the location for $|\omega|_{max}$.
For RUN C, from the very beginning until the collapsing time
($t \approx 0.88$), the location of $|\omega|_{max}$ is
changing continuously and smoothly without any noticeable jumping.
This is quite different to the sudden jumping of
``+'' symbol in RUN A at $t \approx 3.2$.
The $|\nabla \theta|_{max}$ is located around $x \approx 4$
in the beginning (until
$t \approx 0.16$, see ``$*$'' in Fig. \ref{fig:Czoom}),
right on the head of the rising bubble (in this sense, it
behaves like RUN A).

The ``*'' jumps to $x \approx 5$ at $t \approx 0.18$,
roughly at the same time when the ``eye-like'' vortex begins to form.
Afterwards, the ``*'' symbol moves quite continuously towards the
location of $|\omega|_{max}$. After $t=0.7$, the
locations of $|\omega|_{max}$ and $|\nabla \theta|_{max}$
are very close to each other. In the meanwhile, the filament
connecting the cap front and the ``eye'' in the temperature field
becomes thinner and thinner, which breaks down after
$t = 0.87$ due to insufficient resolutions.

For the three low resolutions ($1024^2$, $2048^2$ and $4096^2$), the time evolutions of the three time independent
values $T_2(t)$, $T_4(t)$ and $\theta_{max}(t)$ in RUN C (Fig. \ref{fig:Cerror}) are quite similar to the results in
RUN A (Figs. \ref{fig:Aerror}). Because shorter simulation time is needed, the $4096^2$ result of RUN C is less
affected by the round-off error. As a result, the $4096^2$ curve is always below those of two lower resolutions, see
Fig. \ref{fig:Cerror}(c). The truncation error is the main source of the error for these three low resolutions.

Furthermore, if we we only look at errors of $T_2$ and $T_4$,
the $6144^2$ results show much better conservation property than
that of the three coarser grids (Figs. \ref{fig:Cerror}(a) and (b)).
The $\theta_{max}$ error, which is more easily affected by the round-off
error, shows no advantage of finer grids (Figs. \ref{fig:Cerror}(c)):
for $0\le t\le 0.6$ the error of the $6144^2$ resolutions is at the same
level as that of the $4096^2$ resolution. However, when $t > 0.6$,
the advantage of the $6144^2$ resolution is observed:
the error of $\theta_{max}$ remains around $10^{-6}$ which is one order
lower than that of the $4096^2$ result.
The round-off error of $6144^2$ can roughly
match the truncation error.
This indicates that it seems necessary to employ quadruple
precision instead of the double precision if extremely finer
resolutions (say $10^5$ in each dimension) are used.

Again, we investigate the time evolution of the distance
between the locations of $|\omega|_{max}$ and
$|\nabla \theta|_{max}$
under the same resolution. Unlike the RUN A case (see
Fig. \ref{fig:Adis}), we obtain a very clear physical picture (Fig.
\ref{fig:Cdis}). Around $t = 0.2$, the distance experiences a sudden drop,
and then decays slowly.  The $1024^2$ curve has
an early dramatic increase at $t \approx 0.65$ due to
under-resolution. The
results of the $4096^2$ and $6144^2$ resolutions are quite
satisfactory since the distance remains small. The finest resolution shows no
obvious advantage here. For RUN A, the first-time jump of ``+''
(Figs. \ref{fig:Azoomv}) is followed by the
increasing distance between the locations of
$|\omega|_{max}$ and $|\nabla \theta|_{max}$ (see Fig. \ref{fig:Adis}).
For RUN C, however, it seems that the singularity
forming mechanism is the main driving source, which leads to
the decreasing distance between the locations for
$|\omega|_{max}$ and $|\nabla\theta|_{max}$ after the first jump of ``+''.

Now we can analyze the singularity forming mechanism of RUN C.
We focus on the filament connecting the cap front and
the right ``eye-like'' vortex (see, for example, the temperature
contour plot at $t=0.8$ in Figs. \ref{fig:Czoom}). The
filament corresponds to the hotter region of the fluid field,
which will rise due to the buoyancy without any other forces.
On the other hand, the forming ``eye'' tries to absorb the fluid around it.
These two mechanisms fight with each other, which makes the filament
thinner and thinner until singularity appears. For RUN A,
besides these two mechanisms, there is a third one joining the
competition, namely, the symmetry with respect to $x = \pi$.
The third mechanism is not physical, but the numerical scheme
tends to preserve it throughout the simulations. With three
mechanisms working simultaneously, it is difficult to know
when and where the singularity will be formed, although the time
evolutions of $|\omega|_{max}$ and $|\nabla \theta|_{max}$ give some
hints on solution blow-up. In contrast, the locations of $|\omega|_{max}$ and
$|\nabla \theta|_{max}$ in RUN C are getting closer and closer
when the solutions become singular.

We now think it is safe to carry out some
singularity analysis for RUN C. Figs. \ref{fig:Csing}(a) and (c) show the time
evolutions for $|\omega|_{max}$ and $|\nabla \theta|_{max}$ in
different resolutions; the inverse of the maximum values are plotted
in Figs. \ref{fig:Csing}(b)(dnd ).
In the following, we will pay our attention to the $4096^2$ and $6144^2$
curves, because these two lines seem to be overlapping until around $t=0.88$.

For the $4096^2$ run, there are some very small vorticity structures which
begin to appear at $t=0.84$ in the lower part of the
smooth outer layer where the maximum $|\omega|$ turns up.
For $6144^2$ run, this happens at $ t \geq
0.87$. After this critical time, the filter we adopted in the codes
removes more and more energy from the system. Consequently, the global
average values like $T_2$ are greatly affected. As a result,
our numerical results after $ t \geq 0.87$ may not
be reliable. Actually, it is observed that the $|\omega|_{max}$ and $|\nabla
\theta|_{max}$ experience a drop-down after $t \simeq 0.9$. This
indicates that the filter prevents the occurrence of
the blowup of the maximum vorticity, which is similar to the
viscosity effect in high Reynolds number simulations (e.g.
~\cite{bor94}).  For the $1024^2$ and $2048^2$ runs, however,
the drop-downs appear later than $t \simeq 0.88$ and the
blow-up time $T_c$ is also delayed.

Only the sample maximum values before $t=0.86$ in $6144^2$ run
will be used in the singularity analysis because before this critical
time convergence between the $4096^2$ and $6144^2$ results is observed.
With a statistical weighted least square fitting for the data
after $t = 0.6$, we conclude that the growth of $|\omega|_{max}$
and $|\nabla \theta|_{max}$ for RUN C terminates in a
finite time with $|\omega|_{max} \sim {(T_c-t)}^{-1.12}$, and crudely
$|\nabla \theta|_{max} \sim {(T_c -t)}^{-2.38}$ with $T_c = 0.91$.

There seems to have a trend of singularity also for the $2048^2$ run,
and the $1024^2$ result indicates no blow-up. Moreover, the
collapsing times for higher resolutions are earlier than
those of lower ones because the filter removes more energy with
low resolutions, retarding the singularity forming mechanism.

It should be noticed that the peak vorticity we obtained
is the value modified directly by the filter, and the highest peak
of the $\delta$-like function (see the peak near ``+''
in Fig. \ref{fig:C3dvor} when $t=0.86$) is reduced significantly.
On the other hand, $|\nabla \theta|_{max}$ is less affected by
the filter because it is the temperature field being
filtered (not the $\nabla \theta$ field). No place in the temperature
field is really close to a $\delta$-function.
Hence, the temperature filed is less affected by the filter
than the vorticity field. From this point of view, the $|\nabla
\theta|_{max}$ curve is more accurate than the $|\omega|_{max}$ one.
If there is a blow-up in the simulation, the
$|\nabla \theta|_{max}$ curve will show a stronger divergent
tendency than the $|\omega|_{max}$ one (Figs.
\ref{fig:Csing}).

We have not performed the simulation on grids finer than $6144^2$,
but we can predict some results from the present
computations. It is well known that when a delta function is
approximated by a finite number of Fourier modes, each doubling of
the resolutions will cause doubling of the maximum
value and $2^{n+1}$ times the maximum values of the $n^{th}$-order
space derivative (see the analysis in Appendix A of ~\cite{bor94}).
In the final state of our simulation (Fig.
\ref{fig:Czoom} (t=0.86)), the cut-line at $y = 2.72$
through the out-layer of the ``eye'' of the vorticity field looks
very similar to a delta function. Therefore,
when finer and finer resolutions are used the peak vorticity and
temperature gradient are getting larger and larger. From this
point of view, further 2D Boussinesq simulations with
even higher resolutions will support our singularity
prediction although it seems impossible for any code to reach
the same $T_c$ if the current filtering relevant scheme is used.

\section{Discussions and conclusions}

There have been extensive discussions
on the 3D Euler singularity with viscous simulations by using the
Taylor-Green vortex and high symmetry flows
(the most intensive one, in our opinion is ~\cite{bor94}). As the
Reynolds number is increased, the amplitudes of the maximum vorticity,
skewness, and flatness increase, and the peaks are
attained at earlier times. This is quite similar to what happens to our
simulation with the filter. Our simulation also
reaches an earlier ``blow-up'' when higher resolutions are adopted.
These tendencies give hints that for the pure
inviscid case (no viscosity or filter), all these quantities may blow up.

A natural question is to discuss what is the influence with symmetry
assumptions. We tend to believe that for the
Taylor-Green, high-symmetry Kida and axisymmetric flow, all
the symmetry constraints are unstable, and in the absence of
symmetries, the flow will escape from the singularity formation
direction. Moreover, singularity formations depend strongly on
how the initial conditions are set up. The numerical method chosen has
also a strong impact on the results. For example, finite
difference simulations that do not make use of any symmetry produce no
singularity indication~\cite{pum90} , whereas
Fourier-Chebyshev ~\cite{ker93} simulations with
symmetry constraints suggest singular trends.

To sum up, we follow the track of ~\cite{e94}, and end up with a more developed result. The disagreement between the
locations of $|\omega|_{max}$ and $|\nabla \theta|_{max}$ suggests that the symmetric initial data proposed in
\cite{e94} may not be appropriate (see Fig. \ref{fig:Adis}). To fix this problem, a new initial condition is proposed
in this work, which enables us to make some fine grid simulations.

There are several points that make us believe that
there is a singularity in RUN C:
\begin{itemize}
\item Our simulations show that the time evolution curves
of $|\omega|_{max}$ and $|\nabla \theta|_{max}$ become
steeper when the grids are refined.
\item The distance between $|\omega|_{max}$ and $|\nabla \theta|_{max}$ on the
$6144^2$ and $4096^2$ grid shows a convergent behavior near $T_c$.
\item The distance between the locations of $|\omega|_{max}$
and $|\nabla \theta|_{max}$ is small even up to the predicated blow-up time.
\end{itemize}

\section*{Acknowledgments}
We would like to thank Prof. Linbo Zhang for the support of using the \emph{Lenovo} Deepcom 1800 parallel computer in
the Institute of Computational Mathematics of the Chinese Academy of Sciences. This work was supported by National NSF
of China (G10502054 and G10476032). ZY thanks Professors Ruo Li, Zhenhuan Teng, Jie Shen, Wenrui Hu and Qi Kang for
useful discussions. TT thanks the support from the International Research Team on Complex System of the Chinese Academy
of Sciences and from the Hong Kong Research Grant Council.

\section*{APPENDIX: THE EFFECT OF FILTER ON ROUND-OFF ERROR}

The influence of round-off errors on computation has been observed and noticed
for a long time, probably starting from the appearance of the
digital computer. The round-off error starts to occur at the very beginning
of all numerical simulations, and can be
amplified by the time integration procedure. A comprehensive study on
how machine precision can affect the dynamic
simulation is carried out in ~\cite{kra86}. Later, in
vortex sheet roll-up simulations, a practical result about these
effects with two kinds of precisions (7 and 14-digit arithmetic)
is presented \cite{kra87}. In the following, we will try to
give a brief analysis about the round-off error in our simulations.

\begin{figure*}
\begin{minipage}[c]{.65 \linewidth}
\scalebox{1}[1]{\includegraphics[width=\linewidth]{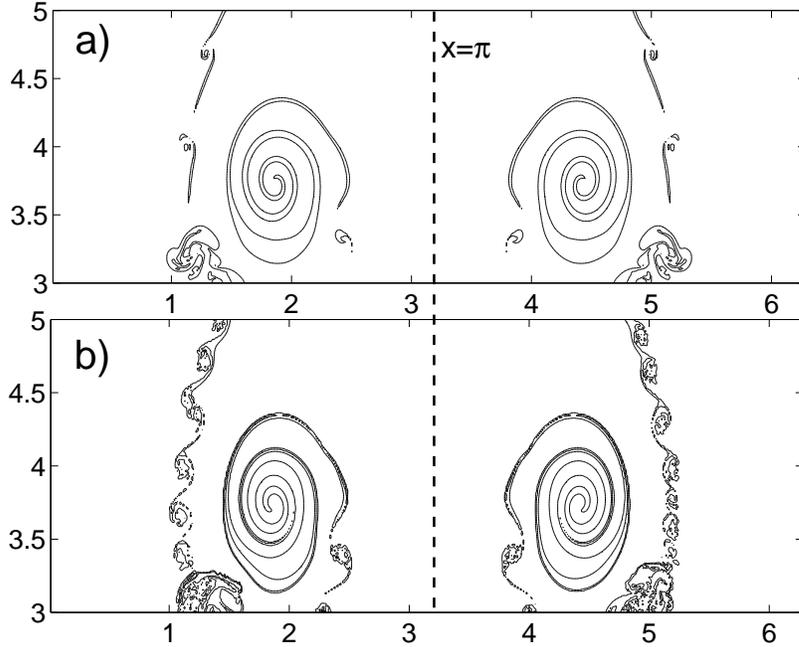}}
\end{minipage}
\begin{minipage}[c]{.25 \linewidth}
\caption{\label{fig:Around} RUN A: Contour plots of temperature and vorticity at different times with the resolution of
$4096^2$.}
\end{minipage}
\end{figure*}

There are three ways to accumulate round-off errors
in our simulations:

\begin{enumerate}
\item
 Fast Fourier Transform (FFT) can control the initial round off error very well. Through the result in
~\cite{tas01,sch96}, we can roughly guess that the error is amplified about 100 times for current resolutions ($1024^2$
$-$ $4096^2$), that is, the error will be in order of
$10^{-13}$ if the double precision ($\varepsilon = 10^{-15}$) is
adopted. The $4096^2$ grid will have only slightly larger round-off error
than that of $1024^2$ grid (only a factor of $
{log 4096}/{log 1024} \approx 1.2$, based on the
worst possibility given in~\cite{tas01}).
\item The filtering technique we adopted can significantly amplify
the round-off error. According to ~\cite{can00}, the filter will
amplify the round-off error by a factor of ${(\sum c_{k}^2)}^{\frac{1}{2}}$.
Here, $c_k$ is the factor added onto the
Fourier coefficients by the filter. Actually, most of $c_k$ are 1,
so roughly speaking, the round-off error will
be enlarged by $N$ times for a $N^2$ simulation. This means our $1024^2$
run will have a $10^{-10}$ error for each time
step, and the error in $4096^2$ runs is four times larger.
\item Remember that the analysis above only happens within each time step,
and the real round-off error can be even
larger in a dynamic run. The time step of $4096^2$ run is  1/4 of
the $1024^2$ run according to the CFL condition,
which means $4096^2$ needs four times as many time steps as $1024^2$.
\end{enumerate}

As a whole, for the same simulation, the round-off error in $4096^2$ run will be $1.2 \times 4 \times 4 = 19.2$ times
larger than that in $1024^2$ run. This difference is already big enough
to break the initial symmetry maintained by the
numerical solver. Fig. \ref{fig:Around}(a) is the contour plot of RUN A at
$t = 4.0$ with the $1024^2$ grid, which has
less details and level than the $4096^2$ grid (Fig. \ref{fig:Around}(b)). However, Fig. \ref{fig:Around}(a) preserves
the symmetry respect to $x = \pi$ very well, while Fig. \ref{fig:Around}(b) has a different number of vortices on the
left and right sides of the picture. In the main text, a discussion of the round-off error is also carried out together
with Figs. \ref{fig:Aerror} and \ref{fig:Cerror}, focusing on the round-off error amplified by the time evolution.

Figs. \ref{fig:Around} are results after the simulation becomes under resolved. Our $4096^2$ simulation of RUN A is
very symmetric respect to $x = \pi$ before $t=3.5$. The round-off error in $6144^2$ will be about 30 times larger than
that of the $1024^2$ run. The round-off error is not negligible comparing with truncation error for $6144^2$ run (Fig.
\ref{fig:Cerror}(c)), it seems a must to adopt quadruple precision on a finer grid.

\end{document}